\documentclass[a4paper,12pt]{article}

\usepackage{amssymb,amsmath,amsfonts,amsthm,mathtools}
\usepackage{eurosym,geometry,ulem,graphicx,caption,color,setspace,sectsty,comment,footmisc,pdflscape,subcaption,array}
\usepackage[inline]{enumitem}
\usepackage{natbib}
\usepackage{hyperref}

\theoremstyle{plain}
\newtheorem{thm}{Theorem}
\newtheorem{lem}{Lemma}
\newtheorem{cor}{Corollary}
\newtheorem{prop}{Proposition}
\newtheorem{fact}{Fact}

\theoremstyle{definition}
\newtheorem{dfn}{Definition}
\newtheorem{rmk}{Remark}
\newtheorem{emp}{Example}
\newtheorem*{prf}{Proof}

\newtheorem*{prfthm2}{Proof of Theorem \ref{thm2}}

\newtheorem*{prfprop3}{Proof of Proposition \ref{prop: equivalent NOM}}

\newcolumntype{L}[1]{>{\raggedright\let\newline\\arraybackslash\hspace{0pt}}m{#1}}
\newcolumntype{C}[1]{>{\centering\let\newline\\arraybackslash\hspace{0pt}}m{#1}}
\newcolumntype{R}[1]{>{\raggedleft\let\newline\\arraybackslash\hspace{0pt}}m{#1}}

\newcommand{\domain}{\mathcal{R}^{n}}
\newcommand{\eqdomain}{\tilde{\mathcal{R}}^{n}}

\DeclareMathOperator*{\argmax}{arg\,max}

\begin{document}

\title{Egalitarian-equivalent and strategy-proof mechanisms in homogeneous multi-object allocation problems\thanks{This work was supported by JST ERATO (JPMJER2301) and the Japan Society for the Promotion of Science (25K16592).
}}

\author{Hinata Kurashita\thanks{Department of Industrial Engineering and Economics, Graduate School of Engineering, Institute of Science Tokyo, 2-12-1 Ookayama, Meguro-ku, Tokyo 152-8552, Japan. e-mail: kurashita.h.aa@m.titech.ac.jp.} ,
Ryosuke Sakai\thanks{School of Economics and Management, University of Hyogo, 8-2-1, Gakuennishi-machi, Nishi-ku, Kobe, Hyogo 651-2197, Japan. e-mail: ryosuke\_sakai@em.u-hyogo.ac.jp}.}

\date{\today}

\maketitle

\begin{abstract}
    We study the problem of allocating homogeneous indivisible objects among agents with money.  In particular, we investigate the relationship between \textit{egalitarian-equivalence} (Pazner and Schmeidler, 1978), as a fairness concept, and \textit{efficiency} under agents’ incentive constraints.
    As a first result, we characterize the class of mechanisms that satisfy \textit{egalitarian-equivalence}, \textit{strategy-proofness}, \textit{individual rationality}, and \textit{no subsidy}. 
    Our characterization reveals a strong tension between \textit{egalitarian-equivalence} and \textit{efficiency}: under these properties, objects are allocated only in limited cases.
    To address this limitation, we replace \textit{strategy-proofness} with the weaker incentive property \textit{non-obvious manipulability} (Troyan and Morrill, 2020).
    We show that this relaxation allows us to design mechanisms that achieve \textit{efficiency} while still ensuring \textit{egalitarian-equivalence}.
    Furthermore, under \textit{efficiency}, we identify the agent-welfare-optimal mechanism  in the characterized class.

\end{abstract}






\doublespacing
\setcounter{page}{1}
\pagenumbering{arabic}

\section{Introduction} \label{sec:introduction}
    We study the problem of allocating homogeneous indivisible objects among agents with money. 
    Governments frequently encounter such situations in contexts like distributing public housing units, granting access to public facilities, or assigning licenses for specific activities.
    While efficiency is an important goal in these processes, fairness is also a crucial consideration, especially when resources are scarce and the needs of agents vary.

    In this study, we focus on an auction environment with unit-demand agents.
    Within this framework, we investigate the implications of a well-known fairness concept, \textit{egalitarian-equivalence} (\citet{pazner1978egalitarian}), and explore its relevance to efficiency and other concepts of fairness.

    We study the problem of allocating $m$ homogeneous indivisible objects among agents with monetary transfers. 
    Each agent has unit demand and preferences over \textit{bundles} consisting of an object assignment and a monetary transfer. 
    Preferences are not necessarily quasi-linear, which allows us to accommodate income effects and soft budget constraints. 
    We refer to the set of all admissible preferences as the \textit{domain}. 
    An agent's \textit{valuation} is her willingness to pay for one object.
    An \textit{object allocation} specifies who receives the objects, and an \textit{allocation} specifies both the object allocation and the monetary transfer of each agent. A \textit{mechanism} is a mapping from preference profiles to allocations.
    \textit{Egalitarian-equivalence} requires that for each preference profile, there exists a reference bundle such that each agent is indifferent between their bundle under the mechanism and the reference bundle. 
    \textit{Strategy-proofness} requires that no agent can achieve a more preferred bundle by reporting a false preference instead of their true preference. 
    \textit{Individual rationality} requires that each agent finds their bundle at least as preferred as the bundle where no objects are received and no monetary transfers are made. 
    \textit{No subsidy} requires that agents make non-negative payments.

    A \textit{$\tau$-augmented minimum uniform price (MUP) mechanism} is a mechanism that allocates objects to agents selected by a winner selection function. 
    Each agent who receives an object pays an amount equal to the $(m+1)$-th highest valuation, while each agent who receives no object makes no payment. 
    Here, a \textit{winner selection function} specifies, for each preference profile, which agents receive objects, subject to the following restrictions: if some agents are selected, then all agents with valuations weakly below the $(m+1)$-th highest valuation have the same valuation, and the number of selected agents does not exceed the number of objects.

    However, a $\tau$-augmented MUP mechanism does not satisfy \textit{egalitarian-equivalence} and \textit{strategy-proofness}. 
    Thus, we introduce two properties on the winner selection function, called \textit{potential winner solidarity} and \textit{uncompromisingness}.
    We show that \textit{potential winner solidarity} is necessary and sufficient for a $\tau$-augmented MUP mechanism to satisfy \textit{egalitarian-equivalence} (Proposition \ref{prop:potential winner solidarity}), and that \textit{uncompromisingness} is necessary and sufficient for it to satisfy \textit{strategy-proofness} (Proposition \ref{prop:uncompromising}).
    
    As a first result, we establish that, on a rich preference domain, a mechanism satisfies  \textit{egalitarian-equivalence}, \textit{strategy-proofness}, \textit{individual rationality}, and \textit{no subsidy} if and only if it is a $\tau$-augmented MUP mechanism whose winner selection function satisfies \textit{potential winner solidarity} and \textit{uncompromisingness} (Theorem \ref{thm1}). 
    This result implies that no objects are allocated in many cases. 
    In other words, no mechanisms satisfy \textit{efficiency}, \textit{egalitarian-equivalence}, \textit{strategy-proofness}, \textit{individual rationality}, and \textit{no subsidy} when the number of objects is fewer than the number of agents minus
    one (Corollary \ref{cor1}). 
    In contrast, when the number of objects equals the number of agents minus one, there exists a mechanism satisfying both \textit{efficiency} and \textit{egalitarian-equivalence} (Corollary \ref{cor2}).

    Motivated by the impossibility established by Corollary \ref{cor1}, we then ask whether weakening the incentive constraints can mitigate the tension. 
    Theorem \ref{thm2} characterizes the class of mechanisms satisfying \textit{egalitarian-equivalence}, \textit
    {efficiency}, \textit{individual rationality}, and \textit{no subsidy}. 
    Such mechanisms must satisfy the following property. If all agents with valuations weakly below the $(m+1)$-th highest valuation have the same valuation, then the allocation must coincide with that of either the Vickrey mechanism or the pay-as-bid mechanism. 
    Otherwise, the allocation must coincide with that of the pay-as-bid mechanism.
    Here, the Vickrey mechanism is a mechanism that allocates the objects so as to maximize the sum of the agents' valuations. Each agent who receives an object pays the $(m+1)$-th highest valuation, while each agent who receives no object makes no payment.
    The pay-as-bid mechanism is a mechanism that allocates the objects so as to maximize the sum of the agents' valuations. Each agent who receives an object pays their own valuation, while each agent who receives no object makes no payment.
    
    We then identify a subclass of this class that satisfies \textit{non-obvious manipulability}, a weak incentive property introduced by \citet{troyan2020obvious} (Theorem \ref{thm3}).
    This possibility, however, is highly limited and depends on delicate circumstances. 
    More specifically, it relies on the possibility that agents may report a preference profile in which all valuations ranked $(m+1)$-th and below are equal to zero.
    Hence, our analysis highlights a limitation of relaxing incentive constraints as a way of alleviating the tension between \textit{egalitarian-equivalence} and \textit{efficiency}.

    The rest of the paper is organized as follows.
    Section \ref{sec:literature} reviews the related literature. 
    Section \ref{sec:model} sets up the model. Section \ref{sec:mechanism} introduces a $\tau$-augmented MUP mechanism. 
    Section \ref{sec:results} provides the main result. 
    Section \ref{sec:achieving efficiency} investigates whether the tension between \textit{egalitarian-equivalence} and \textit{efficiency} can be alleviated by relaxing the incentive constraints.
    Section \ref{sec:conclusion} concludes. 
    All proofs are in the Appendix.\\

\section{Related literature}\label{sec:literature}
    \subsection{Fair allocation}
        This paper focuses on the theory of fair allocation. 
        In this field, two key fairness concepts are widely discussed: \textit{envy-freeness} and \textit{egalitarian-equivalence.}\footnote{For a comprehensive survey, refer to  \citet{thomson2011fair}.} 
        Within the context of the object allocation problem with money, while \textit{envy-freeness} has been extensively studied by many researchers (e.g., \citet{ohseto2006characterizations,papai2003groves, tadenuma1995games}), relatively few studies have examined \textit{egalitarian-equivalence}.

        \citet{ohseto2004implementing}, similar to this paper, studies the problem of allocating  homogeneous indivisible objects in an environment where agents have single-unit demand and quasi-linear preferences, and shows that no mechanism satisfies
        \textit{egalitarian-equivalence}, \textit{strategy-proofness} and \textit{budget balance} on restricted preference domains.\footnote{The intersection of the sets of valuations of each agent contains at least three elements.}
        \citet{shinozaki2022egalitarian} studies the problem of allocating  heterogeneous and indivisible objects in an environment where agents have single-unit demand and non-quasi-linear preferences, and characterizes mechanisms that satisfy \textit{egalitarian-equivalence}, \textit{strategy-proofness}, \textit{individual rationality} and \textit{no subsidy for losers}. The characterized mechanisms result in no allocation of objects in many cases. \citet{yengin2012egalitarian,yengin2017no} considers  the problem of allocating  heterogeneous and indivisible objects in an environment where agents have multi-unit demand and quasi-linear preferences, and identifies the class of mechanisms that satisfy \textit{egalitarian-equivalence}, \textit{strategy-proofness} and \textit{efficiency}. 
       \citet{chun2014egalitarian} investigate the implications of \textit{egalitarian-equivalence} in the context of queueing problems and characterize the class of mechanisms that satisfies \textit{egalitarian-equivalence}, \textit{strategy-proofness} and \textit{queue efficiency}. 

        Unlike these studies, we consider the problem of allocating homogeneous indivisible objects in an environment that is closer to real-world auctions, where some objects remain unallocated and no subsidies are provided to agents.
        Our negative result, which shows that objects are not allocated in most cases, is similar to the findings of  \citet{shinozaki2022egalitarian}. 
        However, our approach to this negative result is different. While \citet{shinozaki2022egalitarian} addresses this problem by relaxing \textit{efficiency}, we explore a different approach by weakening the incentive constraints.

    \subsection{Non-obvious manipulability}
        \citet{troyan2020obvious} propose \textit{non-obvious manipulability} and study the allocation problem of homogeneous objects with multi-demand and quasi-linear preferences. 
        They show that while a pay-as-bid mechanism does not satisfy \textit{non-obvious manipulability}, a uniform-price mechanism does.
        \citet{shinozaki2025non} studies the problem of assigning packages of objects with utilities  that exhibit income effects or face hard budget constraints. 
        Since no mechanism satisfies \textit{strategy-proofness}, \textit{efficiency}, \textit{individual rationality}, and \textit{no subsidy} in that environment, he derives a necessary and sufficient condition for mechanisms that satisfy \textit{non-obvious manipulability}, which is weaker than \textit{strategy-proofness}, under \textit{efficiency}, \textit{individual rationality}, and \textit{no subsidy}.

        Unlike these studies,  we study the problem of allocating homogeneous objects with unit-demand and non-quasi-linear preferences, focusing on the relationship between fairness and efficiency under \textit{non-obvious manipulability}.

        Our study shares the same motivation as that of \citet{psomas2022fair}: overcoming the tension between fairness and efficiency by weakening \textit{strategy-proofness} to \textit{non-obvious manipulability}.\footnote{\citet{psomas2022fair} conduct the analysis using \textit{envy-free up to one item} as a notion  of fairness.}
         While they achieve this in the context of an allocation problem without money, we show that a similar possibility arises in an allocation problem with money.
        Our result (Theorem \ref{thm3}) reinforces the claim that \textit{non-obvious manipulability} enables the flexible design of mechanisms that satisfy both fairness and efficiency.

        \sloppy
        In recent years, \textit{non-obvious manipulability}  has been explored in various models (e.g., \citet{arribillaga2024obvious} and \citet{aziz2021obvious} in a voting model, \citet{ortega2022obvious} in a cake-cutting model, \citet{psomas2022fair} and \citet{troyan2024non} in an
        assignment model without money, \citet{arribillaga2025obvious} in a matching model with contracts, and \citet{archbold2023nona,archbold2023nonb} in an allocation model with money).

        \subsection{Object allocation problems with money when agents have non-quasi-linear preferences}
        There is a large literature on object allocation problems with money when agents have non-quasi-linear preferences. 
        Here, we focus on the environment of this paper, where homogeneous objects are allocated to unit-demand agents. \citet{saitoh2008vickrey} and \citet{sakai2008second} are the first papers to characterize the Vickrey mechanism in this environment by using \textit{efficiency}, \textit{strategy-proofness}, and other mild conditions. 
        In particular,  \citet{sakai2008second} shows that a similar characterization is possible even if \textit{efficiency} is replaced by \textit{envy-freeness}. 
        In addition, there are studies that characterize the Vickrey mechanism using other fairness properties, such as \textit{anonymity in welfare} (\citet{ashlagi2012characterizing}) and \textit{weak envy-freeness for equals} (\citet{adachi2014equity}; \citet{sakai2008second}). This paper complements these studies by analyzing mechanisms that satisfy the fairness property of \textit{egalitarian-equivalence}.
\section{Model} \label{sec:model}
    There are $n$ agents and $m$ homogeneous indivisible objects, where $n>m$. 
    Let $N=\{1,\ldots,n\}$ denote the set of agents. 
    Each agent consumes at most one object. 
    We denote receiving the object and not receiving the object by $1$ and $0$, respectively. 
    A \textbf{consumption bundle} of agent $i\in N$ is $z_i=(x_i,t_i)\in\{0,1\}\times\mathbb{R}$, where $x_i$ is the consumption of the object and $t_i$ is the monetary transfer for agent $i$. 
    Let $\mathbf{0}=(0,0)\in\{0,1\}\times\mathbb{R}$. 
    An \textbf{object allocation} is an $n$-tuple $x=(x_i)_{i\in N}\in \{0,1\}^n$ such that $\sum_{i\in N}x_i\le m$.\footnote{We allow for the possibility that some objects remain unallocated.} 
    Let $X$ denote the set of object allocations. 
    An \textbf{allocation} is an $n$-tuple $z=(z_i)_{i\in N}=(x_i,t_i)_{i\in N}\in\left(\{0,1\}\times\mathbb{R}\right)^n$ such that $(x_i)_{i\in N}\in X$. 
    Let $Z$ denote the set of allocations. 
    We abuse notation and write $z=(x,t)=\mathbf{0}$ if for each $i\in N$, $(x_i,t_i)=\mathbf{0}$. 

    Let $R_i$ be a complete and transitive preference relation of  agent $i\in N$ on $\{0,1\}\times\mathbb{R}$. 
    Let $P_i$ and $I_i$ be the strict and indifferent relations of $R_i$, respectively. We assume that each preference relation satisfies the following conditions:
    \begin{description}
        \item[Money monotonicity.] For each $x_i\in\{0,1\}$ and each $t_i,t'_i\in\mathbb{R}$ with $t_i<t'_i$, $(x_i,t_i)\mathbin{P_i}(x_i,t'_i)$. 
        \item[Weak object desirability.] For each $t_i\in \mathbb{R}$, $(1,t_i)\mathbin{R_i} (0,t_i)$.
        \item[Finiteness.] For each $(x_i,t_i) \in \{0,1\}\times\mathbb{R}$ and each $x'_i\in \{0,1\}$, there is $t'_i \in \mathbb{R}$ such that $(x_i, t_i) \mathbin{I_i} (x'_i, t'_i)$.
    \end{description}

    Let $\mathcal{R}$ denote the set of preference relations. 
    We call $\mathcal{R}$ a \textbf{domain}. 
     Given $R_i \in \mathcal{R}$, $v(R_i) \in \mathbb{R}_+$ is a \textbf{valuation of $R_i$} if $\big(1,v(R_i)\big)\mathbin{I_i} (0,0)$. 
     Given $R_i\in\mathcal{R}$ and $t_i\in\mathbb{R}$, $c_i\big((1,t_i);R_i\big)\in\mathbb{R}$ is a \textbf{compensating valuation for $R_i$ from $(1,t_i)$} if $(1,t_i)\mathbin{I_i}\Big(0,t_i-c_i\big((1,t_i);R_i\big)\Big)$.
    A \textbf{preference profile} is an $n$-tuple $R=(R_i)_{i\in N}\in \domain$. 
    Given $i, j\in N$, let $R_{-i}=(R_j)_{j\in N\setminus\{i\}}$ and let
    $R_{-\{i,j\}}=(R_k)_{k\in N\setminus\{i,j\}}$. 
    For each $k\in\{1,\ldots,n\}$, let $V^k(R) \in \mathbb{R}_+$ be the $k$-th highest valuation for $R \in \domain$, that is, $V^1(R) \ge\cdots\ge V^{k-1}(R)\ge V^k(R)\ge V^{k+1}(R)\ge\cdots \ge V^n(R)$.

    A \textbf{(direct) mechanism on $\domain$} is a function $f\colon \domain\to Z$.
    Given a mechanism $f$ and a preference profile $R\in \domain$, we denote the consumption bundle of agent $i$ for $R$ under $f$ as $f_i(R)=\big(x_i(R),t_i(R)\big)$. 

    Now, we introduce properties for mechanisms.

    Given a mechanism $f$, a preference profile $R\in \domain$ and a bundle $z_0=(x_0,t_0) \in\{0,1\}\times\mathbb{R}$, if $f_i(R)\mathbin{I_i}
    z_0$ for each $i \in N$, then we call the bundle $z_0$ a \textbf{reference bundle for $R$ (under $f$)}.
    The first property, proposed by \citet{pazner1978egalitarian}, is a significant concept of fairness in the literature. 
    It requires that for each preference profile, there is a reference bundle such that each agent is indifferent between their consumption bundle under a mechanism and the reference bundle.
        \begin{description}
            \item[Egalitarian-equivalence.] For each $R\in\domain$, there is a reference bundle $z_0\in \{0,1\}\times\mathbb{R}$ for $R$ under $f$.
        \end{description}
    
    Next, we introduce an incentive property for a mechanism. 
    The second property requires that no agent can benefit by misreporting their true preference. 
        \begin{description}
            \item[Strategy-proofness.] For each $R\in \domain$, each $i\in N$ and each $R'_i\in \mathcal{R}$, $f_i(R)\mathbin{R_i} f_i(R'_i,R_{-i})$.
        \end{description}

            The third property is the (Pareto) efficiency in our model.
                \begin{description}
                    \item[Efficiency.]  For each $R\in\domain$, there is no $z=(x,t)$ such that 
                    \begin{itemize}
                        \item[(i)] for each $i\in N$, $z_i\mathbin{R_i}f_i(R)$,
                        \item[(ii)]  for some $j\in N$, $z_j\mathbin{P_j}f_j(R)$, and
                        \item[(iii)] $\sum_{i\in N}t_i\ge \sum_{i\in N}t_i(R)$
                    \end{itemize}
                \end{description}
                
            We next introduce a voluntary participation condition.
            It requires each agent finds their consumption bundle at least as preferred as receiving nothing with no monetary transfer.
                \begin{description}
                    \item[Individual rationality.] For each $R\in \domain$ and each $i\in N$, $f_i(R)\mathbin{R_i} \mathbf{0}$.
                \end{description}
    
            The last property requires that agents make non-negative monetary transfers.
                \begin{description}
                    \item[No subsidy.] For each $R\in \domain$ and each $i\in N$,  $t_i(R)\ge 0$.
                \end{description}

\section{The $\tau$-augmented minimum uniform price mechanism}\label{sec:mechanism}
    In this section, to state the first result, we propose a \textit{$\tau$-augmented minimum uniform price (MUP) mechanism}. 

    First, we introduce a winner selection function.
    Since the mechanism we consider does not require \textit{efficiency}, we define a selection function that identifies the agents who receive the objects.
    
    To define a winner selection function, we introduce convenient notations. 
    Let $\eqdomain\coloneqq\{R\in\domain\colon V^{m+1}(R)=V^{m+2}(R)=\cdots=V^{n-1}(R)=V^{n}(R)\}$ denote the set of preference profiles such that the valuations ranked $(m+1)$-th or lower are all equal. 
        \begin{dfn}\label{dfn: winner selection function}
            A function $\tau\colon \domain\to 2^N$ is a \textbf{winner selection function} if for each $R\in \domain$, 
                \begin{enumerate*}
                    \item[(i)] if $R\notin\eqdomain$, then  $\tau(R)=\emptyset$, and 
                    \item[(ii)] $|\tau(R)|\le m$.\footnotemark{}
                \end{enumerate*}
        \end{dfn}
        \footnotetext{We denote by $2^A$ the power set of a set $A$, and by $|A|$ the cardinality of a set $A$.}
    Condition (i) requires that if a valuation profile is not in $\eqdomain$, no agents are selected by the winner selection function. 
    Condition (ii) means  that the number of agents selected by the winner selection function does not exceed the number of objects.

    We now introduce a $\tau$-augmented MUP mechanism.
        \begin{dfn}\label{dfn4}
                Let $\tau$ be a winner selection function. A mechanism $f$ on $\domain$ is \textbf{$\tau$-augmented minimum uniform price (MUP) mechanism} if, for each $R\in\domain$ and each $i\in N$,
                \begin{equation*}
                    f_i(R)=
                    \begin{cases}
                     \big(1,V^{m+1}(R)\big) &\text{if } i\in\tau(R),\\
                        \mathbf{0} &\text{if }i\notin\tau(R).
                    \end{cases}
                \end{equation*}
        \end{dfn}
    Note that since $R\notin \eqdomain$ holds in most cases, by the condition (i) in the definition of a winner selection function (Definition \ref{dfn: winner selection function}), all agents receive no object and make no payment.

    The next two examples show that, depending on the winner selection function $\tau$, a $\tau$-augmented MUP mechanism may violate \textit{egalitarian-equivalence} and \textit{strategy-proofness}, respectively.

    \begin{emp}\label{ex:egalitarian-equivalence}
    ($\tau$-augmented MUP mechanisms violate \textit{egalitarian-equivalence}).
    Let $N=\{1,2,3\}$, and suppose that there are two objects. 
    Let $\tau^1$ be a winner selection function such that if $v(R_1)=11$, $v(R_2)=10$, and $v(R_3)=5$, then $\tau^1(R)=\{1\}$.
    Let $f$ be a $\tau^1$-augmented minimum uniform price mechanism. 
    By $f_1(R)=(1,5)$ and $v(R_1)=11$, it must be $z_0=(1,5)$. However, since $z_0=(1,5)\mathbin{P_2}(1,10)\mathbin{I_2}\mathbf{0}=f_2(R)$, $z_0$ is not a reference bundle for $R$, which shows that $f$ violates \textit{egalitarian-equivalence}. 
    \end{emp}

    \begin{emp}\label{ex:strategy-proofness}
    ($\tau$-augmented MUP mechanisms violate \textit{strategy-proofness}).
    Let $N=\{1,2,3\}$, and suppose that there are two objects. 
    Let $\tau^2$ be a winner selection function such that if $v(R_1)=v(R_2)=10$ and $v(R_3)=5$, then $\tau^2(R)=\emptyset$, and if $v(R'_2)=5$, then $\tau^2(R'_{2},R_{-2})=\{2,3\}$.
    Let $f$ be a $\tau^2$-augmented minimum uniform price mechanism. Then, 
    \begin{equation*}
        f_2(R'_2,R_{-2})=(1,5)\mathbin{P_2}(1,10)\mathbin{I_2}(0,0)=f_2(R_2,R_{-2}),
    \end{equation*}
    which shows that $f$ violates \textit{strategy-proofness}.
    \end{emp}
    
    Given Examples \ref{ex:egalitarian-equivalence} and \ref{ex:strategy-proofness}, we impose two properties on the winner selection function  to ensure that  a $\tau$-augmented MUP mechanism satisfies \textit{egalitarian-equivalence} and \textit{strategy-proofness}.

    The first is \textit{potential winner solidarity}. 
    This requires that if there is a potential winner who receives an object, then all potential winners also receive the object, where an agent is called a potential winner if her valuation is strictly higher than the $(m+1)$-th highest valuation. 
    Given $R\in\domain$, an agent $i$ is a \textbf{potential winner for $R$} if $v(R_i)>V^{m+1}(R)$, and let $PW(R)$ denote the set of all potential winners for $R$. 

    \begin{dfn}\label{dfn:potential winner solidarity}
        A winner selection function $\tau$ satisfies \textbf{potential winner solidarity} if for each $R\in\domain$, $PW(R)\cap \tau(R)\ne\emptyset$ implies $PW(R)\subseteq\tau(R)$.
    \end{dfn} 

    The following proposition states a necessary and sufficient condition for a $\tau$-augmented MUP mechanism to satisfy \textit{egalitarian-equivalence}.

    \begin{prop}\label{prop:potential winner solidarity}
        Let $\tau$ be a winner selection function.
        A $\tau$-augmented MUP mechanism satisfies \textit{egalitarian-equivalence} if and only if the winner selection function  satisfies \textit{potential winner solidarity}. 
    \end{prop}

    The second is \textit{uncompromisingness},  which requires an agent selected by the winner selection function to remain selected even if the agent reports another valuation higher than the $(m+1)$-th highest valuation.
    \footnote{\citet{border1983straightforward} introduce \textit{uncompromisingness} in the context of a voting model, while \citet{serizawa1999strategy} employs it  in the context of a public goods provision.  
    In the context of the allocation problem of divisible commodities, \citet{morimoto2013characterization} apply this concept, and \citet{sakai2025fair} first introduced it in the context of object allocation with money.}

    \begin{dfn}\label{dfn:uncompromising}
        A winner selection function $\tau$ satisfies \textbf{uncompromisingness} if for each $R\in\domain$, each $i\in\tau(R)$, and each $R'_i\in \mathcal{R}$ such that $v(R'_i)> V^{m+1}(R)$,
        $i\in\tau(R'_i,R_{-i})$.
    \end{dfn}

    The following proposition states a necessary and sufficient condition for a $\tau$-augmented MUP mechanism to satisfy \textit{strategy-proofness}.

    \begin{prop}\label{prop:uncompromising}
        Let $\tau$ be a winner selection function.
        A $\tau$-augmented MUP mechanism satisfies \textit{strategy-proofness} if and only if the winner selection function  satisfies \textit{uncompromisingness}.  
    \end{prop}

    To avoid redundancy, whenever no confusion arises, we refer to a $\tau$-augmented minimum uniform price mechanism whose winner selection function satisfies \textit{potential winner solidarity} and \textit{uncompromisingness} simply as a \textbf{$\tau$-augmented MUP mechanism}.

\section{Results} \label{sec:results}
\subsection{Richness of the domain}\label{subsec:richness}
    In our results, we need the following richness condition of the domain. 
    \begin{dfn}
        A domain $\mathcal{R}$ is \textbf{rich} if for each $t_i,t'_i>0$ with $t_i<t'_i$, there is $R_i\in\mathcal{R}$ such that $t_i<v(R_i)<t'_i$ and $(0,-t'_i+t_i)\mathbin{P_i}(1,t_i)$. 
    \end{dfn}

    Many economically plausible domains satisfy this richness condition. 
    \begin{emp} 
    The following domains are all rich.
        \begin{itemize}
            \item The class of all preferences satisfying \textit{money monotonicity}, \textit{weak object desirability}, and \textit{finiteness} is rich. 
            \item A preference $R_i$ is \textbf{quasi-linear} if, for each $t_i\in\mathbb{R}$, $c_i\big((1,t_i);R_i\big)=v(R_i)$. 
            We note that, if a preference $R_i$ is quasi-linear, then it is represented by a utility function $u_i\colon\{0,1\}\times\mathbb{R}\to\mathbb{R}$ such that, for each $(x_i,t_i)\in\{0,1\}\times\mathbb{R}$, $u_i(x_i,t_i)=v(R_i)\cdot x_i-t_i$.
            The class of all quasi-linear preferences is rich.
            \item A preference $R_i$ exhibits \textbf{positive income effects}  if for each $t_i,t'_i\in\mathbb{R}$ with $t_i<t'_i$, $c_i\big((1,t_i);R_i\big)>c_i\big((1,t'_i);R_i\big)$. 
            The class of all preferences exhibiting positive income effects is rich. 
            \item A preference $R_i$ exhibits \textbf{non-negative income effects}  if for each $t_i,t'_i\in\mathbb{R}$ with $t_i<t'_i$, $c_i\big((1,t_i);R_i\big)\ge c_i\big((1,t'_i);R_i\big)$. 
            The class of all preferences exhibiting non-negative income effects is rich. 
            \item A preference $R_i$ exhibits \textbf{negative income effects}  if for each $t_i,t'_i\in\mathbb{R}$ with $t_i<t'_i$, $c_i\big((1,t_i);R_i\big)<c_i\big((1,t'_i);R_i\big)$. 
            The class of all preferences exhibiting negative income effects is rich. 
            \item A preference $R_i$ exhibits \textbf{non-positive income effects}  if for each $t_i,t'_i\in\mathbb{R}$ with $t_i<t'_i$, $c_i\big((1,t_i);R_i\big)\le c_i\big((1,t'_i);R_i\big)$. 
            The class of all preferences exhibiting non-positive income effects is rich. 
            \item Given an interest rate $r\in\mathbb{R}_+$, a preference $R_i$ faces a \textbf{soft budget constraint at $r$} if  there exist $w_i\in\mathbb{R}_+$ and  $b_i\in\mathbb{R}_{++}\cup\{\infty\}$ such that for each pair $(x_i,t_i),(x'_i,t'_i)\in\{0,1\}\times\mathbb{R}$, $(x_i,t_i)\mathbin{R_i}(x'_i,t'_i)$ if and only if $w_i\cdot x_i-p(t_i,b_i,r)\ge w_i\cdot x'_i-p(t'_i,b_i,r)$, where for each $t_i\in\mathbb{R}$,
            \begin{equation*}
                p(t_i,b_i,r)=
                \begin{cases}
                     t_i & \text{if $t_i\le b_i$,}\\
                    b_i+(1+r)(t_i-b_i)& \text{if $t_i> b_i$.}
                \end{cases}
            \end{equation*}
            Given an interest rate $r\in\mathbb{R}_+$, the class of all preferences that face a soft budget constraint at $r$ is rich.
        \end{itemize}
    \end{emp}
    \subsection{Characterization}
        As a first result, we characterize a $\tau$-augmented MUP mechanism by \textit{egalitarian-equivalence}, \textit{strategy-proofness}, \textit{individual rationality}, and \textit{no subsidy}.\footnote{This result remains valid if \textit{no subsidy} is replaced by \textit{no subsidy for losers}, which is defined as follows: for each $R\in \domain$ and each $i\in N$, if $x_i(R)=0$, then $t_i(R)\ge0$.}
        
            \begin{thm}\label{thm1}
                Let $\mathcal{R}$ be rich.
                A mechanism $f$ on $\domain$ satisfies \textit{egalitarian-equivalence}, \textit{strategy-proofness}, \textit{individual rationality}, and \textit{no subsidy} if and only if it is a $\tau$-augmented minimum uniform price mechanism whose winner selection function satisfies uncompromisingness and potential winner solidarity. 
            \end{thm}
        Theorem \ref{thm1} presents a negative result, showing that under \textit{strategy-proofness}, \textit{individual rationality} and \textit{no subsidy}, the requirement of \textit{egalitarian-equivalence} implies that, in most cases, neither object allocation nor monetary transfers occur.\footnote{When $n\ge m$, if each agent has a strictly positive valuation, only the mechanism that gives each agent the object for free satisfies \textit{efficiency}, \textit{egalitarian-equivalence}, \textit{strategy-proofness}, \textit{individual rationality}, and \textit{no subsidy}. However, if there exists an agent who does not have a strictly positive valuation, then a mechanism that does not allocate the object to that agent can also satisfy these properties.}
        Objects can be allocated only when the valuations ranked $(m+1)$-th or lower are all equal.
        Hence, we have the following corollary.
            \begin{cor}\label{cor1}
                Assume $m<n-1$. No mechanism on $\domain$ satisfies efficiency,  egalitarian-equivalence, strategy-proofness, individual rationality, and no subsidy. 
            \end{cor}
       When $m=n-1$, the condition defining $\eqdomain$ is vacuous.
        Then, the well-known Vickrey mechanism satisfies \textit{efficiency} in addition to these four properties.
    \begin{dfn}
        A mechanism $f=(x,t)$ on $\domain$ is  a \textbf{Vickrey mechanism} if for each $R\in \domain$ and each $i\in N$, 
        \begin{equation*}
            x(R)\in\argmax_{x\in X}\left\{\sum_{i\in N}v(R_i)\cdot x_i\right\},\quad
            t_i(R)=
                \begin{cases}
                    V^{m+1}(R) &\text{if $x_i(R)=1$,}\\
                    0  &\text{if $x_i(R)=0$.}
                \end{cases}
        \end{equation*}
    \end{dfn}
    We denote by $Z^{\mathrm{V}}(R)$ the set of allocations for $R$ under a Vickrey mechanism. 
    
    Now, we obtain the following corollary.

            \begin{cor}\label{cor2}
                (\citet{saitoh2008vickrey, sakai2008second}). Assume $m=n-1$. A Vickrey mechanism is the only  mechanism that satisfies \textit{efficiency}, \textit{egalitarian-equivalence}, \textit{strategy-proofness}, \textit{individual rationality}, and \textit{no subsidy}.
            \end{cor}

    \subsection{Independence}
        In this subsection, we examine the independence of the properties in Theorem \ref{thm1}. 
        Let $\mathcal{R}$ be a domain satisfying richness. 
        \begin{emp}
            (Dropping \textit{egalitarian-equivalence}).
                Assume $f$ is a $\tau$-augmented MUP mechanism whose winner selection function satisfies \textit{uncompromisingness}, but not \textit{potential winner solidarity} on $\mathcal{R}^n$. 
                Then, by Proposition \ref{prop:potential winner solidarity}, $f$ satisfies \textit{strategy-proofness}, \textit{individual rationality}, and \textit{no subsidy}, but not \textit{egalitarian-equivalence}.
        \end{emp}
        
         \begin{emp}
            (Dropping \textit{strategy-proofness}).  Assume $f$ is a $\tau$-augmented MUP mechanism whose winner selection function satisfies \textit{potential winner solidarity}, but not \textit{uncompromisingness} on $\mathcal{R}^n$. 
            Then, by Proposition \ref{prop:uncompromising}, $f$ satisfies \textit{egalitarian-equivalence}, \textit{individual rationality}, and \textit{no subsidy}, but not \textit{strategy-proofness}.
        \end{emp}
        
         \begin{emp}
            (Dropping \textit{individual rationality}). 
            Assume that $f$ is a mechanism defined on $\mathcal{R}^n$ such that, for each $R\in\mathcal{R}^n$ and each $i\in N$, $f_i(R)=(0,p)$, where $p>0$. 
            Then, $f$ satisfies \textit{egalitarian-equivalence}, \textit{strategy-proofness}, and  \textit{no subsidy}, but not \textit{individual rationality}.
        \end{emp}
        
         \begin{emp}
            (Dropping \textit{no subsidy}).
            Assume that $f$ is a mechanism defined on $\mathcal{R}^n$ such that, for each $R\in\mathcal{R}^n$ and each $i\in N$, $f_i(R)=(0,s)$, where $s<0$. 
            Then, $f$ satisfies \textit{egalitarian-equivalence}, \textit{strategy-proofness}, and \textit{individual rationality}, but not \textit{no subsidy}.
        \end{emp}
        
    \subsection{Relationship with other fairness properties}
        In this subsection, we discuss the relationships between \textit{egalitarian-equivalence} and the other properties of fairness: \textit{envy-freeness} and \textit{anonymity in welfare},\footnote{\textit{Egalitarian-equivalence} \textit{envy-freeness}, and \textit{anonymity in welfare} are logically independent of each other in many models. } under \textit{strategy-proofness}, \textit{individual rationality}, and \textit{no subsidy}. 
        \subsubsection{Envy-freeness}
        \textit{Envy-freeness}, introduced by \citet{foley1967resource}, is another central concept of fairness. 
        It requires that no agent prefers another agent's consumption bundle to their own.
        \begin{description}
            \item[Envy-freeness.] For each $R\in \domain$ and each $i,j \in N$, $f_i(R) \mathbin{R_i} f_j(R)$.
        \end{description}

        Not every mechanism in the class characterized in Theorem \ref{thm1} satisfies \textit{envy-freeness}. 
        The following example shows that some $\tau$-augmented MUP mechanisms that do not satisfy \textit{envy-freeness}.
        \begin{emp}\label{notsatisfyef}
            Let $N=\{1,2,3\}$ and suppose that there are two objects. Let $f$ be a $\tau$-augmented MUP mechanism on $\domain$ whose winner selection function satisfies the following condition: if $v(R_1)=v(R_2)=10$ and $v(R_3)=5$, then $\tau(R)=\{3\}$. 
            Then $f_3(R)=(1,5)\mathbin{P_1}(1,10)\mathbin{I_1}(0,0)=f_1(R)$, which violates \textit{envy-freeness}. 
        \end{emp}

        \subsubsection{Anonymity in welfare}

        \textit{Anonymity in welfare} requires that if the preferences of two agents are exchanged, then their utility levels are also exchanged. 
        In the context of object allocation problems with money, \citet{ashlagi2012characterizing} was the first to introduce this property.
        \begin{description}
            \item[Anonymity in welfare.] For each $R, R' \in \domain$ and each $i,j \in N$, if $R_i = R'_j$, $R_j = R'_i$,
            and $R_{-\{i,j\}} = R'_{-\{i,j\}}$, then $f_i(R)\mathbin{I_i}f_j(R')$.
        \end{description}
        Not all mechanisms in the class characterized in Theorem \ref{thm1} satisfy \textit{anonymity in welfare}. 
        The following example illustrates some  $\tau$-augmented MUP mechanisms  that do not satisfy \textit{anonymity in welfare}.

        \begin{emp}\label{notsatisfyaiw}

            Let $i\in N$ and $f$ be a $\tau$-augmented MUP mechanism on $\domain$ whose winner selection function  satisfies the following conditions: for each $R\in \domain$, 
            \begin{itemize}
                \item if $v(R_i)>2$ and $v(R_j)=2$ for each $j\in N\setminus\{i\}$, then $\tau(R)=\{i\}$, and 
                \item otherwise $\tau(R)=\emptyset$. 
            \end{itemize}
            Then, $f$ satisfies \textit{egalitarian-equivalence}, \textit{strategy-proofness}, \textit{individual rationality}, and \textit{no subsidy}, but not \textit{anonymity in welfare} and \textit{envy-freeness}. 
        \end{emp}
        The next example illustrates some  $\tau$-augmented MUP mechanisms that satisfy \textit{anonymity in welfare}.
        \begin{emp}\label{satisfyaiw}
             Let $f$ be a mechanism on $\domain$ that satisfies the following conditions: for each $R\in \domain$, 
            \begin{itemize}
                \item if $R\in\eqdomain$, then $f(R)\in Z^{\mathrm{V}}(R)$, and 
                \item if $R\notin\eqdomain$, then $f(R)=\mathbf{0}$.
            \end{itemize}
            Then, $f$ satisfies \textit{anonymity in welfare}, \textit{egalitarian-equivalence}, \textit{strategy-proofness}, \textit{individual rationality}, and  \textit{no subsidy}.
            Moreover, $f$ satisfies \textit{envy-freeness}.
        \end{emp}

\section{Relaxing strategy-proofness}\label{sec:achieving efficiency}
    
    In this section, given the impossibility established in Corollary \ref{cor1}, we explore whether the tension between \textit{egalitarian-equivalence} and \textit{efficiency} can be mitigated by relaxing \textit{strategy-proofness}.

        \subsection{Achieving efficiency}
        As a starting point for the analysis, we first abstract from incentive constraints and identify the class of mechanisms satisfying \textit{egalitarian-equivalence}, \textit{efficiency}, \textit{individual rationality}, and \textit{no subsidy}. 

        To state Theorem \ref{thm2}, we introduce a pay-as-bid mechanism.
        \begin{dfn}\label{dfn: pay-as-bid}
         A mechanism $f=(x,t)$ on $\domain$ is a \textbf{pay-as-bid mechanism} if
             for each $R\in \domain$ and each $i\in N$, 
                \begin{equation*}
                    x(R)\in\argmax_{x\in X}\left\{\sum_{i\in N}v(R_i)\cdot x_i\right\},\quad
                    t_i(R)=
                        \begin{cases}
                            v(R_i) &\text{if $x_i(R)=1$,}\\
                            0  &\text{if $x_i(R)=0$.}
                        \end{cases}
                \end{equation*}
        \end{dfn}
    We denote by $Z^{\mathrm{P}}(R)$ the set of allocations for $R$ under a pay-as-bid mechanism. 

    The next theorem characterizes the class of mechanisms satisfying \textit{egalitarian-equivalence}, \textit{efficiency}, \textit{individual rationality}, and \textit{no subsidy}.

        \begin{thm}
        \label{thm2}
         Let $\mathcal{R}$ be rich.
         A mechanism $f$ on $\domain$ satisfies \textit{egalitarian-equivalence}, \textit{efficiency}, \textit{individual rationality}, and \textit{no subsidy} if and only if it satisfies the following conditions:  
         \begin{itemize}
            \item[(i)] if $R\in\eqdomain$, then $f(R)\in Z^{\mathrm{V}}(R)\cup Z^{\mathrm{P}}(R)$, and 
             \item[(ii)] if $R\notin \eqdomain$, then $f(R)\in Z^{\mathrm{P}}(R)$.  
         \end{itemize}
    \end{thm}

    An informal explanation is as follows.
    
    First, when $R \notin\eqdomain$, the only possible reference bundle is $\mathbf{0}$. 
    In order for each agent’s utility level under the mechanism to coincide with that of the reference bundle, any agent who receives an object must pay his valuation induced by the reported preference, while any agent who does not receive an object must pay nothing. 
    Moreover, to achieve \textit{efficiency}, the allocation must maximize the sum of agents’ valuations. 
    In other words, when $R \notin \eqdomain$, the outcome must coincide with that of a pay-as-bid mechanism.

    \subsection{Non-obvious manipulability}
    In this subsection, we investigate whether there exists a mechanism satisfying a weaker incentive property within the class of mechanisms characterized in Theorem \ref{thm2} that satisfy \textit{egalitarian-equivalence}, \textit{efficiency}, \textit{individual rationality}, and \textit{no subsidy}. 
    We then show that there exists a subclass satisfying \textit{non-obvious manipulability}, a weak incentive property introduced by \cite{troyan2020obvious}.
 
    Given a mechanism $f$ on $\domain$ and $R_i,R_i'\in\mathcal{R}$, let $z_i^{\sup}(R'_i;R_i)\in\{0,1\}\times\mathbb{R}$ denote the most preferred bundle for $i$ among the possible bundles under $f$ when the true preference is $R_i$ and the reported preference is $R'_i$. Formally, 
    \begin{itemize}
        \item[(i)] for each $R_{-i}\in\mathcal{R}^{n-1}$, $z_i^{\sup}(R'_i;R_i)\mathbin{R_i}f_i(R'_i,R_{-i})$, and
        \item[(ii)] for each $z_i\in\{0,1\}\times\mathbb{R}$, if for each $R_{-i}\in\mathcal{R}^{n-1}$, $z_i\mathbin{R_i}f_i(R'_i,R_{-i})$, then $z_i\mathbin{R_i}z_i^{\sup}(R'_i;R_i)$. 
    \end{itemize}
    Similarly,  let $z_i^{\inf}(R'_i;R_i)\in\{0,1\}\times\mathbb{R}$ denote the least preferred bundle for $i$ among the possible bundles under $f$ when the true preference is $R_i$ and the reported preference is $R'_i$.
     Formally,
    \begin{itemize}
        \item[(i)] for each $R_{-i}\in\mathcal{R}^{n-1}$, $f_i(R'_i,R_{-i})\mathbin{R_i}z_i^{\inf}(R'_i;R_i)$, and
        \item[(ii)] for each $z_i\in\{0,1\}\times\mathbb{R}$, if for each $R_{-i}\in\mathcal{R}^{n-1}$, $f_i(R'_i,R_{-i})\mathbin{R_i}z_i$, then $z_i^{\inf}(R'_i;R_i)\mathbin{R_i}z_i$. 
    \end{itemize}
    Given a mechanism $f$ on $\domain$, and an agent $i\in N$ with a true preference $R_i\in \mathcal{R}$, a false preference $R'_i\in\mathcal{R}$ is an \textit{obvious manipulation} for agent $i$ with $R_i$
    if the best-case bundle when reporting $R'_i$ is preferred to the best-case bundle when reporting $R_i$, or the worst-case bundle when reporting $R'_i$ is preferred to the worst-case bundle when reporting $R_i$.
    Formally, given a mechanism $f$ on $\domain$, and an agent $i\in N$ with a true preference $R_i\in \mathcal{R}$, a false preference $R'_i \in \mathcal{R}$ is an \textbf{obvious manipulation} for agent $i$ with $R_i$ if either 
        \begin{align*}
            z_i^{\sup}(R'_i;R_i)\mathbin{P_i}z_i^{\sup}(R_i;R_i)
            \text{ or }z_i^{\inf}(R'_i;R_i)\mathbin{P_i}z_i^{\inf}(R_i;R_i).\footnotemark{}   
        \end{align*}
            \footnotetext{\citet{troyan2020obvious} define \textit{obvious manipulation} by comparing the maximum and minimum utilities an agent can achieve by misreporting versus reporting truthfully, assuming a model with a finite outcome set. 
            In contrast, the outcome (allocation) set in our model is infinite, which means that maximum or minimum utilities may fail to exist. 
            To address this technical challenge, we instead rely on definitions involving the supremum and infimum, following \citet{ortega2022obvious} and \citet{shinozaki2025non}. 
            Importantly, this refinement does not alter the substance of our main results. 
            Even under the original notion of \textit{non-obvious manipulability} based on maxima and minima, Proposition \ref{prop: equivalent NOM} and Theorem \ref{thm3} continue to hold.}

            \textit{Non-obvious manipulability} requires that there is no obvious manipulation for each agent $i$ with any $R_i$. 
            \begin{description}
                \item[Non-obvious manipulability.] For each $i\in N$ and each $R_i,R'_i\in \mathcal{R}$,  
            \begin{align*}
            z_i^{\sup}(R_i;R_i)\mathbin{R_i}z_i^{\sup}(R'_i;R_i)
            \text{ and }z_i^{\inf}(R_i;R_i)\mathbin{R_i}z_i^{\inf}(R'_i;R_i). 
        \end{align*}
            \end{description}
        Note that \textit{non-obvious manipulability} is weaker than \textit{strategy-proofness}.

         \begin{rmk}\label{rmk: NOM}
                \citet{troyan2020obvious} provide a behavioral characterization of \textit{obvious manipulation}. 
                Given a mechanism $f$ on $\domain$, an agent $i\in N$, and  a preference  $R_i \in \mathcal{R}$, let $\pi_i^f(R_i)=\{z_i\in\{0,1\}\times\mathbb{R}\colon\text{$\exists R_{-i}\in \mathcal{R}^{n-1}$ s.t. $f_i(R_i,R_{-i})=z_i$}\}$ denote the set of bundles possible for agent $i$ under $f$ given $R_i$. 
                Given a mechanism $f$ on $\domain$ and an agent $i\in N$, another mechanism $g$ on $\domain$ is \textbf{$i$-indistinguishable from} $f$ if for each $R_i\in \mathcal{R}$, $\pi_i^f(R_i) = \pi_i^g(R_i)$. 
                Theorem 1 in \citet{troyan2020obvious} shows that when there are at least three agents and a set of valuations is rich in a way \footnote{Their richness condition requires that the cardinality of $\mathcal{R}$ should be greater than or equal to that of $\{0,1\}\times\mathbb{R}$. Our domain, as well as every subclass of the domain introduced in Section \ref{subsec:richness}, satisfies their richness condition.},
                the following holds:
                a report $R'_i$ is an \textit{obvious manipulation} under a mechanism $f$ on $\domain$ at $R_i$ if and only if for each mechanism $g$ on $\domain$ that is $i$-indistinguishable from $f$, $R'_i$ is a profitable manipulation of $g$ at $R_i$. 
                That is, an \textit{obvious manipulation} can be interpreted as a manipulation recognizable even by agents with limited cognitive ability, and \textit{non-obvious manipulability} can be understood as an incentive property that eliminates such manipulations.
            \end{rmk}
    To this end, we first derive an equivalent condition to \textit{non-obvious manipulability} under \textit{efficiency}, \textit{individual rationality}, and \textit{no subsidy}. 
    This condition requires that the best-case bundle under truthful reporting be indifferent to receiving an object for free. 
    This result is a restatement of Theorem 1 in \citet{shinozaki2025non}, adapted to our model.
     \begin{prop}\label{prop: equivalent NOM}
        Let $\mathcal{R}$ be rich.
        Let $f$ be a mechanism on $\domain$ that satisfies efficiency, individual rationality, and no subsidy. Then, $f$ satisfies non-obvious manipulability if and only if for each $i\in N$ and each $R_i\in\mathcal{R}$, 
        \begin{equation*}
            z^{\sup}_i(R_i;R_i)\mathbin{I_i}(1,0). 
        \end{equation*}
    \end{prop}

    Now, within the class characterized in Theorem \ref{thm2}, we identify a subclass that satisfies \textit{non-obvious manipulability}.

     \begin{thm}
        \label{thm3}
         Let $\mathcal{R}$ be rich.
         A mechanism $f$ on $\domain$ satisfies \textit{egalitarian-equivalence}, \textit{efficiency}, \textit{non-obvious manipulability}, \textit{individual rationality}, and \textit{no subsidy} if and only if it satisfies the following conditions:  
         \begin{itemize}
            \item[(i)] if $R\in\eqdomain$, then $f(R)\in Z^{\mathrm{V}}(R)\cup Z^{\mathrm{P}}(R)$,
             \item[(ii)] if $R\notin \eqdomain$, then $f(R)\in Z^{\mathrm{P}}(R)$, and  
             \item[(iii)] for each $i\in N$ and each $R_i\in\mathcal{R}$ with $v(R_i)>0$, $\inf\{t_i(R_i,R_{-i})\colon \exists R_{-i}\in\mathcal{R}^{n-1}, x_i(R_i,R_{-i})=1\}=0$.\footnotemark{}
         \end{itemize}
    \end{thm}
    \footnotetext{By Proposition 5 in \citet{shinozaki2025non}, our result remains unchanged even if we consider implementation in \textit{non-obviously manipulable strategies} by indirect mechanisms rather than direct mechanisms.}

    Within the class characterized in Theorem \ref{thm2}, a mechanism such that agents who receive objects pay their valuations  induced by reported preferences does not satisfy the conditions in Proposition \ref{prop: equivalent NOM}, implying that \textit{non-obvious manipulability} is not satisfied. Therefore, when $R\in\eqdomain$ and the valuations from the $(m+1)$-th and beyond are identical at $0$, it is necessary to apply a Vickrey mechanism. 
    Then, the condition in Proposition \ref{prop: equivalent NOM} is satisfied.

    For other cases when $R \in \eqdomain$, either the reference bundle is $\mathbf{0}$ or $\big(1,V^{m+1}(R)\big)$, 
    so it is necessary to apply either a Vickrey mechanism, or a pay-as-bid mechanism. 

    \begin{rmk}
        The result of Theorem \ref{thm3} is limited and depends on delicate circumstances.  
        Specifically, it relies on the existence of a preference profile in which all valuations ranked $(m+1)$-th and below are equal to zero.
    \end{rmk}

    \subsection{Relationship with other fairness properties}
        In this subsection, we discuss the relationships between \textit{egalitarian-equivalence} and the other properties of fairness under \textit{non-obvious  manipulability}, \textit{efficiency}, \textit{individual rationality}, and \textit{no subsidy}. 
        
        The mechanisms in the class characterized in Theorem \ref{thm3} do not satisfy \textit{envy-freeness} because, when a pay-as-bid mechanism is applied, payments can differ even if agents receive objects. 
        
        The characterized mechanisms do not necessarily satisfy \textit{anonymity in welfare}.
        According to the condition (iii) in Theorem \ref{thm3}, the allocation under a Vickrey mechanism may lead to different utility levels, even when the valuations are exchanged between agents, depending on  valuation profiles. 
        
    \subsection{Welfare}
        In this subsection, in line with the goal of achieving \textit{efficiency} in this section, we propose a mechanism that optimizes  welfare of agents under \textit{egalitarian-equivalence}, \textit{non-obvious  manipulability}, \textit{efficiency}, \textit{individual rationality}, and \textit{no subsidy}. 
        
        First, we denote by $F$ the class of mechanisms that satisfy \textit{egalitarian-equivalence}, \textit{non-obvious manipulability}, \textit{efficiency}, \textit{individual rationality}, and \textit{no subsidy}. 
    
        Next, we define the welfare dominance relationship between mechanisms.

    \begin{dfn}
        A mechanism $f$ on $\domain$ \textbf{welfare dominates} another mechanism $f'$ on $\domain$ if for each $R\in \domain$ and $i\in N$, $f_i(R)\mathbin{R_i} f'_i(R)$.
    \end{dfn}
   A mechanism $f$ is \textbf{agent-welfare-optimal} in a class of mechanisms if it is in the class and welfare dominates every other mechanism in the class. 

   Next, we define a specific mechanism based on the characterization of the class of mechanisms that satisfy \textit{egalitarian-equivalence}, \textit{non-obvious manipulability}, \textit{efficiency}, \textit{individual rationality}, and \textit{no subsidy}, as presented in Theorem \ref{thm3}.

    \begin{dfn}
        A mechanism $f$ on $\domain$ is a $\eqdomain$-Vickrey mechanism if for each $R\in \domain$, 
            \begin{itemize}
                \item if $R\in\eqdomain$, then $f(R)\in Z^{\mathrm{V}}(R)$ and 
                \item if $R\notin \eqdomain$, then $f(R)\in Z^{\mathrm{P}}(R)$. 
            \end{itemize}
    \end{dfn}

    Finally, we identify an agent-welfare-optimal mechanism among the class of mechanisms that satisfy \textit{egalitarian-equivalence}, \textit{non-obvious manipulability}, \textit{efficiency}, \textit{individual rationality}, and \textit{no subsidy}.

     \begin{prop}\label{prop3}
       Let $\mathcal{R}$ be rich.
       Let $F$ be the class of mechanisms on $\domain$ that satisfy \textit{egalitarian-equivalence}, \textit{non-obvious manipulability}, \textit{efficiency}, \textit{individual rationality}, and \textit{no subsidy}. 
       A $\eqdomain$-Vickrey mechanism is the unique agent-welfare-optimal mechanism among $F$.
        \end{prop}

        By Theorem \ref{thm2}, when $R\in\eqdomain$, any \textit{agent-welfare-optimal} mechanism must choose either the allocation of a Vickrey mechanism or that of a pay-as-bid mechanism. 
        Since \textit{agent-welfare-optimality} requires minimizing each agent's payment, the allocation must be the one induced by a Vickrey mechanism.

\section{Conclusion} \label{sec:conclusion}
    
    In this paper, as a first result, we have characterized the class of mechanisms that satisfy \textit{egalitarian-equivalence}, \textit{strategy-proofness}, \textit{individual rationality} and \textit{no subsidy} (Theorem \ref{thm1}).
    This result shows that when the number of objects is fewer than the number of agents minus one, in most cases, agents are unable to receive any objects. 
    This highlights a strong trade-off between \textit{efficiency} and \textit{egalitarian-equivalence} (Corollary \ref{cor1}). 
    In contrast, when the number of objects equals the number of agents minus one, both \textit{efficiency} and \textit{egalitarian-equivalence} can be achieved simultaneously (Corollary \ref{cor2}).

    Given these results, we explore whether the tension between \textit{egalitarian-equivalence} and \textit{efficiency} can be mitigated by weakening incentive constraints. 
    As a second result, we first identify the class of mechanisms that satisfy \textit{egalitarian-equivalence}, \textit{efficiency}, \textit{individual rationality}, and \textit{no subsidy} (Theorem \ref{thm2}), and then find a subclass that also satisfies \textit{non-obvious manipulability} (Theorem \ref{thm3}). 
    However, this possibility is highly limited and depends on delicate circumstances. 
    Hence, our analysis reveals a limitation of relaxing incentive constraints as an approach: weakening incentive constraints does not generally resolve the tension between \textit{egalitarian-equivalence} and \textit{efficiency}.

\section*{Appendix: Proofs} \label{sec:appendix}
\addcontentsline{toc}{section}{Appendix: Proofs}
\renewcommand{\thesubsection}{\Alph{subsection}}
    \subsection{Preliminary lemmas}
        
    In this subsection, we prove the lemmas used to establish the theorems and propositions.

    \begin{lem}\label{lem1}
         Let $f$ be a mechanism on $\domain$ that satisfies individual rationality and no subsidy. Let $i\in N$ and $R\in \domain$.
        If $x_i(R)=0$, then $f_i(R)=\mathbf{0}$.
    \end{lem}
    \begin{prf}
        Suppose $x_i(R)=0$. 
        By \textit{individual rationality}, $\big(0,t_i(R)\big)\mathbin{R_i}\mathbf{0}$.
        By \textit{money monotonicity}, this implies $t_i(R)\le 0$.
        By \textit{no subsidy}, $t_i(R)\ge 0$. 
        Hence, $t_i(R)=0$.
        Therefore, $f_i(R)=\mathbf{0}$. \qed
    \end{prf}

    \begin{lem}\label{lem2}
         Let $f$ be a mechanism on $\domain$ that satisfies individual rationality and no subsidy. 
        Let $R\in \domain$ and $z_0 = (x_0, t_0) \in \{0,1\}\times\mathbb{R}$ be a reference bundle for $R$.
        If $x_0 =0$, then $z_0 =\mathbf{0}$.
    \end{lem}
    \begin{prf}
         Suppose $x_0 = 0$. By $n>m$, there is $i \in N$ such that $x_i(R)= 0$. 
         By Lemma \ref{lem1}, $f_i(R)=\mathbf{0}$. 
         Since $f_i(R)=\mathbf{0}$ and $f_i(R)\mathbin{I_i}z_0=(0,t_0)$, we have $(0,0)\mathbin{I_i}(0,t_0)$. By \textit{money monotonicity}, this is possible only if $t_0=0$. 
         Hence, $z_0=\mathbf{0}$. \qed
    \end{prf}

    \begin{lem}\label{lem3}
        Let $f$ be a mechanism on $\domain$ that satisfies egalitarian-equivalence, individual rationality, and no subsidy. Let $R\in\domain$ and $i\in N$. If $f_i(R)\mathbin{P_i}\mathbf{0}$, then 
        \begin{enumerate*}
            \item[(i)] $x_i(R)=1$ and
            \item[(ii)] the reference bundle for $R$ is $f_i(R)$. 
        \end{enumerate*} 
    \end{lem}

    \begin{prf}
        Assume $f_i(R)\mathbin{P_i}\mathbf{0}$. 
        
        (i) Suppose $x_i(R)=0$. Then, $f_i(R)=\big(0,t_i(R)\big)\mathbin{P_i}\mathbf{0}$. 
        By \textit{money monotonicity}, $t_i(R)<0$, which contradicts \textit{no subsidy}. 

        (ii) Let $z_0=(x_0,t_0)$ be the reference bundle for $R$. 
        If $x_0=0$, then by Lemma \ref{lem2}, $z_0=\mathbf{0}$.
        If $x_0=1$, then since $f_i(R)\mathbin{I_i}z_0$ and, by part (i), $f_i(R)=\big(1,t_i(R)\big)$, \textit{money monotonicity} implies $z_0=f_i(R)$.
        Thus, $z_0\in\{\mathbf{0},f_i(R)\}$. Since $f_i(R)\mathbin{P_i}\mathbf{0}$,  we must have $z_0=f_i(R)$.\qed
    \end{prf}

    \begin{lem}\label{lem4}
        Let $\mathcal{R}$ be rich. Let $f$ be a mechanism on $\domain$ that satisfies egalitarian-equivalence, strategy-proofness, individual rationality, and no subsidy. 
        Let $R\in \domain$ and $i\in N$. If $x_i(R)=1$, then $t_i(R)=V^{m+1}(R)$.
    \end{lem}

    \begin{prf} 
        First, to prove this lemma, we show the following fact (Lemma 8 in \cite{morimoto2015strategy}). 
        \begin{fact}\label{fact: SP}
        Let $f$ be a mechanism on $\domain$ that satisfies strategy-proofness, individual rationality, and no subsidy. 
        Let $R\in\domain$, $i\in N$, and $R'_i\in\mathcal{R}$ be such that $x_i(R)=1$ and $t_i(R)<v(R'_i)$.
        Then, $f_i(R'_i,R_{-i})=f_i(R)$.
        \end{fact}
        \begin{prf}
            Suppose $f_i(R'_i,R_{-i})\ne f_i(R)$. 
            Assume $x_i(R'_i,R_{-i})=0$.
            Then, by Lemma \ref{lem1}, $f_i(R'_i,R_{-i})=\mathbf{0}$.
            Hence, by $x_i(R)=1$ and $t_i(R)<v(R'_i)$, $f_i(R)\mathbin{P'_i}f_i(R'_i,R_{-i})$, which contradicts \textit{strategy-proofness}.
            Therefore, we have $x_i(R'_i,R_{-i})=1$. 
            By $f_i(R'_i,R_{-i})\ne f_i(R)$ and $x_i(R)=x_i(R'_i,R_{-i})=1$, $t_i(R'_i,R_{-i})\ne t_i(R)$, and so $f_i(R)\mathbin{P'_i}f_i(R'_i,R_{-i})$ or $f_i(R'_i,R_{-i})\mathbin{P_i}f_i(R)$. This also contradicts \textit{strategy-proofness}.\qed
        \end{prf}
        We now prove Lemma \ref{lem4}.
        Assume $x_i(R)=1$. 
        Suppose $t_i(R)<V^{m+1}(R)$.
        Let $R'_i\in\mathcal{R}$ be such that $t_i(R)<v(R'_i)<V^{m+1}(R)$.
        Such a preference exists by richness.
        By Fact \ref{fact: SP}, $f_i(R'_i,R_{-i})=f_i(R)$. 
        Let $z_0$ be the reference bundle for $(R'_i,R_{-i})$. 
        Since $f_i(R'_i,R_{-i})\mathbin{P'_i}\mathbf{0}$, by Lemma \ref{lem3} (ii) and $f_i(R'_i,R_{-i})=f_i(R)$, we have $z_0=f_i(R)$. 
        Note that by $t_i(R)<V^{m+1}(R)$, $|\{j\in N\setminus\{i\}\colon v(R_j)>t_i(R)\}|\ge m$. 
        Also note that for each $j\in N\setminus\{i\}$ with $v(R_j)>t_i(R)$, $f_j(R'_i,R_{-i})\mathbin{I_j}z_0=f_i(R)\mathbin{P_j}\mathbf{0}$.
        Hence, by Lemma \ref{lem3} (i), for each $j\in N\setminus\{i\}$ with $v(R_j)>t_i(R)$, $x_j(R'_i,R_{-i})=1$. 
        Therefore, $|\{k\in N\colon x_k(R'_i,R_{-i})=1\}|\ge m+1$.
        This implies that $x(R'_i,R_{-i})$ is not an object allocation, which is a contradiction. 
        Hence, we have $t_i(R)\ge V^{m+1}(R)$. 

        Suppose $t_i(R)>V^{m+1}(R)$. 
        Let $R'_i\in\mathcal{R}$ be such that $v(R'_i)>t_i(R)$.
        Such a preference exists by richness.
        By Fact  \ref{fact: SP}, $f_i(R'_i,R_{-i})=f_i(R)$.
        Let $z_0$ be the reference bundle for $(R'_i,R_{-i})$. 
        Since $f_i(R'_i,R_{-i})\mathbin{P'_i}\mathbf{0}$, by Lemma \ref{lem3} (ii) and $f_i(R'_i,R_{-i})=f_i(R)$, we have $z_0=f_i(R)$. 
        Note that by $t_i(R)>V^{m+1}(R)$, there is $j\in N\setminus\{i\}$ such that $v(R_j)\le V^{m+1}(R)<t_i(R)$.    
        Since $v(R_j)\le V^{m+1}(R)<t_i(R)$, we have $\mathbf{0}\mathbin{P_j}\big(1,t_i(R)\big)=z_0$.
        Since $z_0$ is a reference bundle for $(R'_i,R_{-i})$, $f_j(R'_i,R_{-i})\mathbin{I_j}z_0$.
        Hence, $\mathbf{0}\mathbin{P_j}f_j(R'_i,R_{-i})$, which violates \textit{individual rationality}.
        Therefore, we have $t_i(R)=V^{m+1}(R)$.\qed
    \end{prf}

    \begin{lem}\label{lem5}
        Let $f$ be a mechanism on $\domain$ satisfying egalitarian-equivalence, strategy-proofness, individual rationality and no subsidy. 
        Let $R\in \domain$ and $i\in N$. If $x_i(R)=1$, then $v(R_i)\ge V^{m+1}(R)$.
    \end{lem}

    \begin{prf}
           Assume $x_i(R)=1$. By Lemma \ref{lem4} and \textit{individual rationality}, $f_i(R)=\big(1,V^{m+1}(R)\big)\mathbin{R_i} \mathbf{0}\mathbin{I_i}\big(1,v(R_i)\big)$. Thus, by \textit{money monotonicity}, $v(R_i)\ge V^{m+1}(R)$.\qed
    \end{prf}

    \begin{lem}\label{lem:rich,eff,||=m}
             Let $\mathcal{R}$ be rich. Let $f$ be a mechanism on $\domain$ that satisfies efficiency, individual rationality, and no subsidy. For each $R\in\domain$, if for each $i\in N$, $v(R_i)>0$, then $|\{l\in N\colon x_l(R)=1\}|=m$. 
        \end{lem}
        \begin{prf}
            Suppose by contradiction that there is $R\in\domain$ such that 
             $v(R_i)>0$ for each $i\in N$ and $|\{l\in N\colon x_l(R)=1\}|<m$.
             Then, there is $j\in N$ such that $x_j(R)=0$. 
             By \textit{individual rationality}, \textit{no subsidy}, and Lemma \ref{lem1}, $f_j(R)=\mathbf{0}$. 
             Let $z\in Z$ be such that $z_i=(1,0)$ and $z_j=f_j(R)$ for each $j\in N\setminus\{i\}$. 
             Then, $z_i\mathbin{P_i}f_i(R)$, $z_j\mathbin{I_j}f_j(R)$, and $\sum_{i\in N}t_i=\sum_{i\in N}t_i(R)$. 
             Hence, $f$ does not satisfy \textit{efficiency}, which is a contradiction. \qed
        \end{prf}

    \subsection{Proof of Proposition \ref{prop:potential winner solidarity}}
    Let $f$ be a $\tau$-augmented MUP mechanism on $\domain$.
         
        \mbox{}\\
        \textbf{If.}
        Assume that $\tau$ satisfies \textit{potential winner solidarity}. 
        Suppose by contradiction that $f$ does not satisfy \textit{egalitarian-equivalence}.
        Then, there is $R\in\domain$ such that there is no reference bundle for $R$. 
        This implies that there is $i\in N$ such that $f_i(R)\mathbin{P_i}\mathbf{0}$ or $\mathbf{0}\mathbin{P_i}f_i(R)$. 
        By the definition of $f$, $f_i(R)\mathbin{P_i}\mathbf{0}$.  
        Hence, $f_i(R)=\big(1,V^{m+1}(R)\big)$, $\tau(R)\ne\emptyset$, and $i\in\tau(R)\cap PW(R)$. 
        By $\tau(R)\ne\emptyset$ and the condition (i) of a winner selection function, we have $R\in\eqdomain$. 
        Since there is no reference bundle for $R$, there is $j\in N\setminus\{i\}$ such that $f_j(R)\mathbin{P_j}\big(1,V^{m+1}(R)\big)$ or $\big(1,V^{m+1}(R)\big)\mathbin{P_j}f_j(R)$. 
        If $x_j(R)=1$, by the definition of $f$, $f_j(R)=\big(1,V^{m+1}(R)\big)$, which is a contradiction. 
        Thus, $x_j(R)=0$, and by Lemma \ref{lem1}, $f_j(R)=\mathbf{0}$. 
        Hence, we have $\big(1,V^{m+1}(R)\big)\mathbin{P_j}f_j(R)=\mathbf{0}$. 
        Therefore, $j\in PW(R)$ and $j\notin\tau(R)$, which contradicts \textit{potential winner solidarity}. 

        \mbox{}\\
        \textbf{Only if.}
        Assume that $f$ satisfies \textit{egalitarian-equivalence}. 
        Suppose by contradiction that $\tau$ does not satisfy \textit{potential winner solidarity}.
        That is, there is $R\in\domain$ such that $PW(R)\cap\tau(R)\ne\emptyset$ and $PW(R)\nsubseteq\tau(R)$. 
        Let $z_0\in\{0,1\}\times\mathbb{R}$ be the reference bundle for $R$, and let $i,j\in PW(R)$ be such that $i\in\tau(R)$ and $j\notin\tau(R)$. 
        By $i,j\in PW(R)$, we have $v(R_i)> V^{m+1}(R)$ and $v(R_j)> V^{m+1}(R)$.
        Hence, by $i\in\tau(R)$, $f_i(R)=\big(1,V^{m+1}(R)\big)\mathbin{P_i}\mathbf{0}$,
        and so $z_0\in\left\{\big(1,V^{m+1}(R)\big), \big(0,t_i(R)-c_i\big(f_i(R);R_i\big)\big)\right\}$.
        Note that $\big(0,t_i(R)-c_i\big(f_i(R);R_i\big)\big)\ne\mathbf{0}$.
        Moreover, by $j\notin\tau(R)$, $f_j(R)=\mathbf{0}$. 
        Hence $z_0\in\{\big(1,v(R_j)\big),\mathbf{0}\}$. 
        Also note that by $j\in PW(R)$, $\big(1,v(R_j)\big)\ne\big(1,V^{m+1}(R)\big)$.
        Hence, $\left\{\big(1,V^{m+1}(R)\big), \big(0,t_i(R)-c_i\big(f_i(R);R_i\big)\big)\right\}\cap\{\big(1,v(R_j)\big),\mathbf{0}\}=\emptyset$, which means that there is no reference bundle for $R$. 
        This contradicts \textit{egalitarian-equivalence}.\qed 

    \subsection{Proof of Proposition \ref{prop:uncompromising}}
         Let $f$ be a $\tau$-augmented MUP mechanism on $\domain$.
         
        \mbox{}\\
        \textbf{If.}
        Assume $\tau$ satisfies \textit{uncompromisingness}. 
        Let $R\in \domain$, $i\in N$, and $R'_i\in \mathcal{R}$. 

        We prove by cases based on whether objects are allocated or not for each preference profile.
        It is trivial to show that $f$ satisfies \textit{strategy-proofness} when 
            \begin{itemize}
                \item  $x_i(R)=0$ and $x_i(R'_i,R_{-i})=0$,
                \item $x_i(R)=1$ and $x_i(R'_i,R_{-i})=0$, and
                \item $x_i(R)=1$ and $x_i(R'_i,R_{-i})=1$.
            \end{itemize} 

        Assume $x_i(R)=0$ and $x_i(R'_i,R_{-i})=1$.
        Then $f_i(R)=\mathbf{0}$ by the definition of $f$.
        We claim that $v(R_i)\le V^{m+1}(R'_i,R_{-i})$.
        Suppose instead that $v(R_i)>V^{m+1}(R'_i,R_{-i})$.
        Since $i\in\tau(R'_i,R_{-i})$, uncompromisingness applied to the profile $(R'_i,R_{-i})$ and the report $R_i$ implies $i\in\tau(R_i,R_{-i})$, which contradicts $x_i(R)=0$.
        Thus, $v(R_i)\le V^{m+1}(R'_i,R_{-i})$. 
        Since $f_i(R'_i,R_{-i})=\big(1,V^{m+1}(R'_i,R_{-i})\big)$, we have $\mathbf{0}\mathbin{I_i}\big(1,v(R_i)\big)
        \mathbin{R_i}\big(1,V^{m+1}(R'_i,R_{-i})\big)
        =f_i(R'_i,R_{-i})$.
        Hence $i$ cannot profit by reporting $R'_i$.
        
         \mbox{}\\
        \textbf{Only if.}
        Assume $f$ satisfies \textit{strategy-proofness}.
        Suppose by contradiction $\tau$ does not satisfy  \textit{uncompromisingness}. That is, for some $i\in\tau(R)$ and some $R'_i\in\mathcal{R}$ with  $v(R'_i)> V^{m+1}(R)$, $i\notin\tau(R'_i,R_{-i})$. 
        Then $f_i(R'_i,R_{-i})=\mathbf{0}$. 
        However, 
                \begin{equation*}
                    f_i(R) =\big(1,V^{m+1}(R)\big)
                    \mathbin{P'_i}\big(1,v(R'_i)\big)
                    \mathbin{I'_i}\mathbf{0}
                    =f_i(R'_i,R_{-i}),
                \end{equation*}    
        which violates  \textit{strategy-proofness}. Thus, $i\in\tau(R'_i,R_{-i})$.\qed
        
    \subsection{Proof of Theorem \ref{thm1}}
        Let $\mathcal{R}$ be rich. 
        
        \mbox{}\\
        \textbf{If.} Let $\tau$ be a winner selection function that satisfies \textit{potential winner solidarity} and \textit{uncompromisingness}, and let $f$ be a $\tau$-augmented MUP mechanism on $\domain$.
        It is trivial to show that $f$ satisfies \textit{individual rationality} and \textit{no subsidy}, so we omit the proof. 
         By Proposition \ref{prop:potential winner solidarity}, $f$ satisfies \textit{egalitarian-equivalence}.
        By Proposition \ref{prop:uncompromising}, $f$ satisfies \textit{strategy-proofness}.
       
        \mbox{}\\
        \textbf{Only if.}
        Let $f$ be a mechanism on $\domain$ that satisfies \textit{egalitarian-equivalence}, \textit{strategy-proofness}, \textit{individual rationality}, and \textit{no subsidy}.
            The proof is in two steps.\\
        \textsc{Step 1.} We construct a winner selection function. Given $R\in \domain$, let $\tau^f(R)=\{i\in N\colon x_i(R)=1\}$. We show that $\tau^f$ satisfies each condition of a winner selection function (Definition \ref{dfn: winner selection function}). 
        \begin{enumerate}    
            \item[(i)] \textit{For each $R\in \domain$, if $R\notin\eqdomain$, then $\tau^f(R)=\emptyset$.} 

            By contradiction, suppose that there is $R \notin\eqdomain$ such that $\tau^f(R) \ne \emptyset$.
            Let $i\in N$ be such that $x_i(R)=1$.
            By Lemma  \ref{lem5}, $v(R_i)\ge V^{m+1}(R)$. 
             By Lemma \ref{lem4}, $f_i(R)=\big(1,V^{m+1}(R)\big)$.
            Assume there is $R'_i\in\mathcal{R}$ with $v(R'_i)> v(R_i)$.  
            Since $v(R_i)\ge V^{m+1}(R)$, we have $v(R'_i)>V^{m+1}(R)=t_i(R)$.
            By Fact \ref{fact: SP}, $f_i(R'_i,R_{-i})=f_i(R)=\big(1,V^{m+1}(R)\big)$.
            Since $v(R'_i)>V^{m+1}(R)$, we have $f_i(R'_i,R_{-i})=\big(1,V^{m+1}(R)\big)\mathbin{P'_i}\mathbf{0}$.
            Let $z_0$ be a reference bundle for $(R'_i,R_{-i})$.
            By Lemma \ref{lem3}, applied to the profile $(R'_i,R_{-i})$, $z_0=f_i(R'_i,R_{-i})=\big(1,V^{m+1}(R)\big)$.
            Since $R\notin\eqdomain$, there exists $j\in N\setminus\{i\}$ such that $v(R_j)<V^{m+1}(R)$.
            For this agent, $\mathbf{0}\mathbin{I_j}\big(1,v(R_j)\big)\mathbin{P_j}\big(1,V^{m+1}(R)\big)=z_0$.
            Since $z_0$ is a reference bundle for $(R'_i,R_{-i})$,
            $f_j(R'_i,R_{-i})\mathbin{I_j}z_0$.
            Hence, $\mathbf{0}\mathbin{P_j}f_j(R'_i,R_{-i})$, which contradicts \textit{individual rationality}.

            \item[(ii)] \textit{For each $R\in \domain$, $|\tau^f(R)|\le m$.} \\
           Directly follows from the definition of $\tau^f$. 
        \end{enumerate}
        \textsc{Step 2.}  Now, we complete the proof.
        Assume $i\notin\tau(R)$. By the definition of $\tau^f$, $x_i(R)=0$. By Lemma \ref{lem1}, $t_i(R)=0$.
        Thus, $f_i(R)=\mathbf{0}$. 
        Assume $i\in\tau(R)$. By the definition of $\tau^f$, $x_i(R)=1$. 
        By Lemma \ref{lem4}, $t_i(R)=V^{m+1}(R)$.  
        Thus, $f$ is the $\tau$-augmented MUP mechanism. 
        By \textit{strategy-proofness} and Proposition \ref{prop:uncompromising}, the winner selection function $\tau^f$ satisfies \textit{uncompromisingness}. 
        By \textit{egalitarian-equivalence} and Proposition \ref{prop:potential winner solidarity}, the winner selection function $\tau^f$ satisfies \textit{potential winner solidarity}. 
        \qed

    \subsection{Proof of Proposition \ref{prop: equivalent NOM}}
    First, we prove lemmas to prove Proposition \ref{prop: equivalent NOM}.

    \begin{lem}\label{lem:rich,eff,f_i(R)=0}
            Let $\mathcal{R}$ be rich. Let $f$ be a mechanism on $\domain$ that satisfies efficiency, individual rationality, and no subsidy. 
            For each $i\in N$ and each $R_i\in\mathcal{R}$, there is $R_{-i}\in\mathcal{R}^{n-1}$ such that $f_i(R_i,R_{-i})=\mathbf{0}$.
        \end{lem}

        \begin{prf}
            Let $i\in N$ and $R_i\in\mathcal{R}$. 
            By \textit{finiteness}, there is $c\in\mathbb{R}$ such that $c\ge\max_{t_i\in[0,v(R_i)]}c_i\big((1,t_i);R_i\big)$.
            Let $R_{-i}\in\mathcal{R}^{n-1}$ be such that, for each $j\in N\setminus\{i\}$, $v(R_j)>c$. 
            Suppose $x_i(R_i,R_{-i})=1$. 
            Then, there is $j \in N\setminus\{i\}$ such that $x_j(R_i, R_{-i})=0$.
            Let $z=(x,t)\in Z$ such that 
                \begin{itemize}
                    \item[(i)] $z_i=\big(0,t_i(R_i,R_{-i})-c\big)$,
                    \item[(ii)] $z_j=\big(1,v(R_j)\big)$, and 
                    \item[(iii)] $z_k=f_k(R_i,R_{-i})$ for each $k\in N\setminus\{i,j\}$. 
                \end{itemize}
            Note that by the definitions of a compensating valuation and $c$, 
                \begin{equation*}
                    \bigg(0,t_i(R_i,R_{-i})-c_i\Big(\big(1,t_i(R_i,R_{-i})\big);R_i\Big)\bigg)\mathbin{I_i}f_i(R_i,R_{-i})
                \end{equation*}
            and $c\ge c_i\Big(\big(1,t_i(R_i,R_{-i})\big);R_i\Big)$.
            Hence, $z_i=\big(0,t_i(R_i,R_{-i})-c\big)\mathbin{R_i}f_i(R_i,R_{-i})$. 
            By $f_j(R_i,R_{-i})=\mathbf{0}$, we have $z_j\mathbin{I_j}\big(1,v(R_j)\big)$. 
            Therefore, for each $l\in N$, $z_l\mathbin{R_l}f_l(R)$. 
            By $v(R_j)>c$ and $t_j(R_i,R_{-i})=0$,
            \begin{align*}
                \sum_{l\in N}t_l &= 
                t_i(R_i,R_{-i})-c+v(R_j)+\sum_{k\in N\setminus\{i,j\}}t_k(R_i,R_{-i})\\
                &>t_i(R_i,R_{-i})+0+\sum_{k\in N\setminus\{i,j\}}t_k(R_i,R_{-i})\\
                &=\sum_{l\in N}t_l(R_i,R_{-i}).
            \end{align*}
            Hence, $f(R_i,R_{-i})$ is not \textit{efficient} for $(R_i,R_{-i})$, which is a contradiction. \qed
        \end{prf}

        \begin{lem}\label{lem:rich,eff,x_i(R)=1}
            Let $\mathcal{R}$ be rich. Let $f$ be a mechanism on $\domain$ that satisfies efficiency, individual rationality, and no subsidy. For each $i\in N$ and each $R_i\in\mathcal{R}$ with $v(R_i)>0$, there is $R_{-i}\in\mathcal{R}^{n-1}$ such that $x_i(R_i,R_{-i})=1$. 
        \end{lem}

        \begin{prf}
            Let $i\in N$ and $R_i\in\mathcal{R}$ with $v(R_i)>0$. 
            By \textit{richness}, there is $R'_{-i}\in\mathcal{R}^{n-1}$ such that for each $j\in N\setminus\{i\}$, $0<v(R_j)<\frac{v(R_i)}{2}$, and $\left(0,-\frac{v(R_i)}{2}\right)\mathbin{P_j}(1,0)$. 
            Suppose that $x_i(R_i,R'_{-i})=0$. 
            By \textit{individual rationality}, \textit{no subsidy}, and Lemma \ref{lem1}, $f_i(R_i,R'_{-i})=\mathbf{0}$. 
            By \textit{efficiency}, \textit{individual rationality}, \textit{no subsidy}, and Lemma \ref{lem:rich,eff,||=m}, $|\{l\in N\colon x_l(R_i,R'_{-i})=1\}|=m$. 
            Let $j\in N\setminus\{i\}$ be such that $x_j(R_i,R'_{-i})=1$. 
            Then, since $\left(0,-\frac{v(R_i)}{2}\right)\mathbin{P_j}(1,0)$, by \textit{individual rationality} and \textit{no subsidy}, we have $0\le t_j(R_i,R'_{-i})\le v(R_j)<\frac{v(R_i)}{2}$. 
            Let $z\in Z$ be such that $z_i=\big(1,v(R_i)\big)$, $z_j=\left(0,-\frac{v(R_i)}{2}\right)$, and for each $k\in N\setminus\{i,j\}$, $z_k=f_k(R_i,R'_{-i})$. 
            Then, $z_i\mathbin{I_i}f_i(R_i,R'_{-i})$, $z_j\mathbin{P'_j}f_j(R_i,R'_{-i})$, $z_k=f_k(R_i,R'_{-i})$ for each $k\in N\setminus\{i,j\}$, and 
            \begin{align*}
                \sum_{l\in N}t_l &=v(R_i)-\frac{v(R_i)}{2}+\sum_{k\in N\setminus\{i,j\}}t_k(R_i,R'_{-i})\\
                &=\frac{v(R_i)}{2}+\sum_{k\in N\setminus\{i,j\}}t_k(R_i,R'_{-i}) \\
                &> 0+t_j(R_i,R'_{-i}) +\sum_{k\in N\setminus\{i,j\}}t_k(R_i,R'_{-i}) \\
                &= \sum_{l\in N}t_l(R_i,R'_{-i}). 
            \end{align*}
            Hence, $f$ does not satisfy \textit{efficiency}, which is a contradiction. 
            Thus, $x_i(R_i,R'_{-i})=1$.\qed
        \end{prf}

        Next, we prove Proposition \ref{prop: equivalent NOM}.
        \begin{prfprop3}
        
         Let $\mathcal{R}$ be rich. Let $f$ be a mechanism on $\domain$ that satisfies \textit{efficiency}, \textit{individual rationality}, and \textit{no subsidy}.   The proof follows the method of \citet{shinozaki2025non}.

        \mbox{}\\
        \textbf{If.}
        Assume that for each $i\in N$ and each $R_i\in\mathcal{R}$,
        \begin{equation*}
            z_i^{\sup}(R_i;R_i)\mathbin{I_i}(1,0).
        \end{equation*}
        By \textit{no subsidy}, for each $R'_i\in\mathcal{R}$,
            \begin{equation*}
                (1,0)\mathbin{R_i}z_i^{\sup}(R'_i;R_i).
            \end{equation*}
            Since $z_i^{\sup}(R_i;R_i)\mathbin{I_i}(1,0)$, we obtain  
            \begin{equation*}
                z_i^{\sup}(R_i;R_i)\mathbin{R_i}z_i^{\sup}(R'_i;R_i).
            \end{equation*}

        Next, we show that for each $i\in N$ and each $R_i,R'_i\in\mathcal{R}$,
        \begin{equation*}
             z_i^{\inf}(R_i;R_i)\mathbin{R_i}z_i^{\inf}(R'_i;R_i).
        \end{equation*}
        Let $i\in N$ and $R_i,R'_i\in\mathcal{R}$. 
        By \textit{individual rationality},
        \begin{equation*}
            z_i^{\inf}(R_i;R_i)\mathbin{R_i}\mathbf{0}.
        \end{equation*}
        By Lemma \ref{lem:rich,eff,f_i(R)=0}, there is $R_{-i}\in\mathcal{R}^{n-1}$ such that $f_i(R'_i,R_{-i})=\mathbf{0}$. 
        Thus, 
        \begin{equation*}
            \mathbf{0}\mathbin{R_i}z_i^{\inf}(R'_i;R_i).
        \end{equation*}
        Hence, we have
        \begin{equation*}
            z_i^{\inf}(R_i;R_i)\mathbin{R_i}z_i^{\inf}(R'_i;R_i).
        \end{equation*}

        \mbox{}\\
        \textbf{Only if.}
         Assume that $f$ satisfies \textit{non-obvious manipulability}. Suppose by contradiction that there is $i\in N$ with $R_i\in \mathcal{R}$ such that             \begin{equation}
                \label{inequality1}
                (1,0)\mathbin{P_i}z_i^{\sup}(R_i;R_i).\footnotemark{}
            \end{equation}
             Let $\overline{\mathcal{R}}^{n-1}\coloneqq \{R_{-i}\in \mathcal{R}^{n-1}\colon x_i(R_i,R_{-i})=1\}$. 
             By \textit{individual rationality}, \textit{no subsidy}, and Lemma \ref{lem1}, we have
             \begin{equation*}
                 (1,0)\mathbin{P_i} z_i^{\sup}(R_i;R_i)\mathbin{R_i} \mathbf{0},
             \end{equation*}
            which implies $v(R_i)>0$. 
            Therefore, Lemma \ref{lem:rich,eff,x_i(R)=1} implies that $\overline{\mathcal{R}}^{n-1}$ is nonempty.
             By inequality (\ref{inequality1}), $\inf_{R'_{-i}\in \overline{\mathcal{R}}^{n-1}}t_i(R_i,R'_{-i})>0$. Then, there is  $R'_i\in \mathcal{R}$ such that
                \begin{equation*}
                    \inf_{R'_{-i}\in \overline{\mathcal{R}}^{n-1}}t_i(R_i,R'_{-i})>v(R'_i)>0. 
                \end{equation*}
            Let $R_{-i}\in\mathcal{R}^{n-1}$ be such that for each $j \in N \setminus\{i\}$, $v(R_j)=0$. By \textit{efficiency} and $v(R'_i)>0$, $x_i(R'_i,R_{-i})=1$. By \textit{individual rationality}, 
                \begin{equation*}
                    t_i(R'_i,R_{-i})\le v(R'_i)< \inf_{R'_{-i}\in \overline{\mathcal{R}}^{n-1}}t_i(R_i,R'_{-i}).
                \end{equation*}
            Thus, 
            \begin{equation*}
                z_i^{\sup}(R'_i;R_i) \mathbin{R_i}\big(1,t_i(R'_i,R_{-i})\big) 
                \mathbin{P_i} \Big(1,\inf_{R'_{-i}\in \overline{\mathcal{R}}^{n-1}}t_i(R_i,R'_{-i})\Big) 
                =z_i^{\sup}(R_i;R_i).
            \end{equation*}
            which contradicts \textit{non-obvious manipulability}. 
        \qed
\end{prfprop3}
     \subsection{Proof of Theorem \ref{thm2}}
     First, we prove the following lemma to prove Theorem \ref{thm2}.
     \begin{lem}
        \label{lem6}
         Let $\mathcal{R}$ be rich.
         Let $f$ be a mechanism on $\domain$ that satisfies \textit{egalitarian-equivalence}, \textit{efficiency}, \textit{individual rationality}, and \textit{no subsidy}. 
        Let $R\notin\eqdomain$. 
        Then, $f(R)\in Z^{\mathrm{P}}(R)$.
    \end{lem}
    \begin{prf}
        Let $f$ be a mechanism on $\domain$ that satisfies \textit{egalitarian-equivalence}, \textit{efficiency}, \textit{individual rationality}, and \textit{no subsidy}.
     Let $R\notin\eqdomain$.
     Now, we show that a reference bundle $z_0$ for $R$ is $\mathbf{0}$. 
    Suppose $z_0 \ne \mathbf{0}$.
    Then, by Lemma \ref{lem2}, $x_0=1$,
    and so there is $p \in \mathbb{R}$ such that $z_0=(1,p)$. 
    By $R\notin \eqdomain$ and Lemma \ref{lem:rich,eff,||=m}, there are $i,j\in N$ such that $V^{m+1}(R)=v(R_i)>v(R_j)$, $x_i(R) = 0$, and $x_j(R)=0$.
    By Lemma \ref{lem1}, $f_i(R)=\mathbf{0}$ and $f_j(R)=\mathbf{0}$. 
    By  $z_0=(1,p)$, $z_0=(1,p)\mathbin{I_k}f_k(R)=\mathbf{0}\mathbin{I_k}\big(1,v(R_k)\big)$ for each $k\in\{i,j\}$, which implies $v(R_i)=v(R_j)$. 
    This contradicts $v(R_i)>v(R_j)$.
    
    Thus, $z_0=\mathbf{0}$.
    Then, for each $i\in N$ such that $x_i(R)=1$,
        \begin{equation*}
            f_i(R)=\big(1,t_i(R)\big)\mathbin{I_i}z_0=\mathbf{0}\mathbin{I_i}\big(1,v(R_i)\big),
        \end{equation*}
    which implies $t_i(R)=v(R_i)$. 
    For each $j\in N$ such that $x_j(R)=0$, by Lemma \ref{lem1},
        \begin{equation*}
            f_j(R)=\mathbf{0}\mathbin{I_j}z_0.
        \end{equation*}
    Thus, $f(R)\in Z^{\mathrm{P}}(R)$.\qed
    \end{prf}

    Next, we prove Theorem \ref{thm2}.

     \begin{prfthm2}\mbox{}\\
        \textbf{If.} Let $f$ be a mechanism on $\domain$ that satisfies the following conditions:
            \begin{itemize}
            \item[(i)] if $R\in\eqdomain$, then $f(R)\in Z^{\mathrm{V}}(R)\cup Z^{\mathrm{P}}(R)$, and
             \item[(ii)] if $R\notin \eqdomain$, then $f(R)\in Z^{\mathrm{P}}(R)$.
         \end{itemize}

         It is trivial to show that $f$ satisfies \textit{individual rationality} and \textit{no subsidy}, so we omit the proof. We now prove that $f$ satisfies \textit{egalitarian-equivalence} and \textit{efficiency}.

        \mbox{}\\
         \textsc{Egalitarian-equivalence.}
            By the condition (i), if $R\in\eqdomain$, then $f(R)\in Z^{\mathrm{V}}(R)\cup Z^{\mathrm{P}}(R)$. Thus $\big(1,V^{m+1}(R)\big)$ or $\mathbf{0}$ is a reference bundle for $R$. By the condition (ii), if $R\notin \eqdomain$, then $f(R)\in Z^{\mathrm{P}}(R)$. Thus $\mathbf{0}$ is a reference bundle for $R$.

        \mbox{}\\
         \textsc{Efficiency.}
         First, we prove the lemma to prove that $f$ satisfies \textit{efficiency}.
         \begin{lem}\label{lem:eff,equiv}
            Let $f$ be a mechanism on $\domain$ that satisfies \textit{individual rationality} and \textit{no subsidy}.
            Assume that 
            \begin{enumerate}
                \item[(i)] for each $R\in\domain$, 
                    \begin{equation*}
                        x(R)\in\arg\max_{x\in X}\left\{\sum_{i\in N}v(R_i)\cdot x_i\right\}, 
                    \end{equation*}
                    and 
                \item[(ii)]for each $i\in N$ with $x_i(R)=1$, $t_i(R)\ge V^{m+1}(R)$. 
            \end{enumerate} 
            Then, $f$ satisfies efficiency. 
         \end{lem}
         \begin{prf}
             Let $R\in\domain$. 
             Assume that there is $z=(x,t)\in Z$ such that $z_i\mathbin{R_i}f_i(R)$ for each $i\in N$ and $z_j\mathbin{P_j}f_j(R)$ for some $j\in N$. 
             Let $N_1=\{i\in N\colon x_i=1\}$ and $N_2=\{i\in N\colon x_i(R)=1\}$. 
             We show that  $\sum_{i\in N}t_i<\sum_{i\in N}t_i(R)$.
             Note that since for each $i\in N$, $z_i\mathbin{R_i}f_i(R)$, 
             \begin{itemize}
                 \item for each $i\in N_1\cap N_2$, $t_i\le t_i(R)$, 
                 \item for each $i\in N\setminus(N_1\cup N_2)$, by Lemma \ref{lem1}, $t_i\le t_i(R)=0$. 
                 \item for each $i\in N_1\setminus N_2$, by Lemma \ref{lem1} and assumption (i), $t_i(R)=0$ and $t_i\le  V^{m+1}(R)$, and 
                 \item for each $i\in N_2\setminus N_1$, by \textit{individual rationality} and assumption (ii), $t_i\le 0$ and $ t_i(R)\ge V^{m+1}(R)$,
             \end{itemize}
             Hence, 
             \begin{equation*}
                 \sum_{i\in N}t_i \le\sum_{i\in N_1}t_i\le \sum_{i\in N_1\cap N_2}t_i(R)+\sum_{i\in N\setminus(N_1\cup N_2)}0+\sum_{i\in N_1\setminus N_2}V^{m+1}(R)+\sum_{i\in N_2\setminus N_1}0.
             \end{equation*}
             By $z_j\mathbin{P_j}f_j(R)$, if $j\in N_1\cap N_2$, then $t_j<t_j(R)$; if $j\in N_1\setminus N_2$, then $t_j<V^{m+1}(R)$; and if $j\in N_2\setminus N_1$, then $t_j<0$. 
             Therefore,
             \begin{equation*}
                 \sum_{i\in N_1}t_i<\sum_{i\in N_1\cap N_2}t_i(R)+\sum_{i\in N_1\setminus N_2}V^{m+1}(R).
             \end{equation*}
             Moreover, if $V^{m+1}(R)>0$, assumption (i) implies $|N_2|=m$ and $|N_1\setminus N_2|\le|N_2\setminus N_1|$.
             Hence, if $V^{m+1}(R)>0$, then
             \begin{align*}
                 \sum_{i\in N_1}t_i &<\sum_{i\in N_1\cap N_2}t_i(R)+\sum_{i\in N_1\setminus N_2}V^{m+1}(R)\\
                 &< \sum_{i\in N_1\cap N_2}t_i(R)+\sum_{i\in N_2\setminus N_1}V^{m+1}(R)\\
                 &\le \sum_{i\in N}t_i(R). 
             \end{align*}
             If $V^{m+1}(R)=0$, then 
             \begin{align*}
                 \sum_{i\in N_1}t_i &< \sum_{i\in N_1\cap N_2}t_i(R)+\sum_{i\in N_1\setminus N_2}V^{m+1}(R)\\
                 &\le \sum_{i\in N_1\cap N_2}t_i(R)\\
                 &\le \sum_{i\in N}t_i(R). 
             \end{align*}
             Therefore, $\sum_{i\in N}t_i<\sum_{i\in N}t_i(R)$, and so $f$ satisfies \textit{efficiency}. \qed
         \end{prf}
         By the definitions of a Vickrey mechanism and a pay-as-bid mechanism, 
            $f=(x,t)$ satisfies that for each $R\in \domain$,
            \begin{equation*}
                x(R)\in\argmax_{x\in X}\left\{\sum_{i\in N}v(R_i)\cdot x_i\right\},
            \end{equation*}
            and for each $i\in N$ with $x_i(R)=1$, $t_i(R)\ge V^{m+1}(R)$. 
            By Lemma \ref{lem:eff,equiv}, $f$ satisfies \textit{efficiency}.

        \mbox{}\\
        \textbf{Only if.}
            Let $f$ be a mechanism on $\domain$ that satisfies  \textit{egalitarian-equivalence}, \textit{efficiency}, \textit{individual rationality}, and \textit{no subsidy}.  Let $R\in \domain$ and  $i\in N$. 
            By Lemma \ref{lem6}, if $R\notin\eqdomain$, then $f(R)\in Z^{\mathrm{P}}(R)$.

            Assume $R\in\eqdomain$. Under \textit{egalitarian-equivalence}, the reference bundle for $R$ is either $\big(1,V^{m+1}(R)\big)$ or $\mathbf{0}$. By \textit{egalitarian-equivalence} and \textit{efficiency},
            for $i\in N$ such that $x_i(R)=1$,
                \begin{equation*}
                    f_i(R)=\big(1,t_i(R)\big)=\big(1,V^{m+1}(R)\big),
                \end{equation*}
                which implies $t_i(R)=V^{m+1}(R)$, or 
                \begin{equation*}
                     f_i(R)=\big(1,t_i(R)\big)=\big(1,v(R_i)\big)\mathbin{I_i}\mathbf{0},
                \end{equation*}
                which implies $t_i(R)=v(R_i)$.
            By Lemma \ref{lem1}, for $i\in N$ such that $x_i(R)=0$,  $t_i(R)=0$. 
        Thus,  $f(R)\in Z^{\mathrm{V}}(R)\cup Z^{\mathrm{P}}(R)$. \qed
     \end{prfthm2}

     \subsection{Proof of Theorem \ref{thm3}}
     \textbf{If.}
     Let $f$ be a mechanism on $\domain$ that satisfies the following conditions:
            \begin{itemize}
            \item[(i)] if $R\in\eqdomain$, then $f(R)\in Z^{\mathrm{V}}(R)\cup Z^{\mathrm{P}}(R)$,
             \item[(ii)] if $R\notin \eqdomain$, then $f(R)\in Z^{\mathrm{P}}(R)$, and  
             \item[(iii)]  for each $i\in N$ and each $R_i\in\mathcal{R}$ with $v(R_i)>0$, $\inf\{t_i(R_i,R_{-i})\colon \exists R_{-i}\in\mathcal{R}^{n-1}, x_i(R_i,R_{-i})=1\}=0$.
         \end{itemize}
      It is enough to show that $f$ satisfies \textit{non-obvious manipulability}.
        By the condition (iii), $f$ satisfies the condition of Proposition \ref{prop: equivalent NOM}. Thus, $f$ satisfies \textit{non-obvious manipulability}.
        
     \mbox{}\\
        \textbf{Only if.}
         Let $f$ be a mechanism on $\domain$ that satisfies  \textit{egalitarian-equivalence}, \textit{non-obvious manipulability}, \textit{efficiency}, \textit{individual rationality}, and \textit{no subsidy}. 
         Let $R\in \domain$ and  $i\in N$. 
        It is enough to show that condition (iii) holds.
        By Lemma \ref{lem:rich,eff,x_i(R)=1} and Proposition \ref{prop: equivalent NOM}, for each $i\in N$ and each $R_i\in \mathcal{R}$ with $v(R_i)>0$, $\inf\{t_i(R_i,R_{-i})\colon \exists R_{-i}\in\mathcal{R}^{n-1}, x_i(R_i,R_{-i})=1\}=0$.\qed

     \subsection{Proof of Proposition \ref{prop3}}

         Let $f^{\mathrm{V}}$ be a Vickrey mechanism and let $f^{\mathrm{P}}$ be a pay-as-bid mechanism on $\domain$.
         Since a Vickrey mechanism satisfies \textit{individual rationality}, it is straightforward to show that for each $R\in \domain$ and each $i\in N$, 
        \begin{equation}
         \label{inequalityVP}
            f^{\mathrm{V}}_i(R)\mathbin{R_i} f^{\mathrm{P}}_i(R)\mathbin{R_i} \mathbf{0}.
        \end{equation} 
        Let $F$ be the class of mechanisms on $\domain$ that satisfy \textit{egalitarian-equivalence}, \textit{non-obvious manipulability}, \textit{efficiency}, \textit{individual rationality}, and \textit{no subsidy}.
        By Theorem \ref{thm2}, for each $f\in F$, $R\notin\eqdomain$, and $i\in N$, $f_i(R)=\big(1,v(R_i)\big)=f^{\mathrm{P}}_i(R)$. By inequality (\ref{inequalityVP}), for each $f\in F$, $R\in\eqdomain$, and $i\in N$, 
        \begin{equation}
        \label{inequalityVf}
             f^{\mathrm{V}}_i(R)\mathbin{R_i}f_i(R).
        \end{equation}
        Let $f^*\in F$ be a $\eqdomain$-Vickrey mechanism. 
        Then, by inequality (\ref{inequalityVP}) and (\ref{inequalityVf}), for each $f\in F$, $R\in \domain$, and $i\in N$, 
        \begin{equation*}
            f^*_i(R)\mathbin{R_i} f_i(R), 
        \end{equation*}
        which implies $f^*$ is \textit{agent-welfare-optimal} among $F$. 

        Finally, we prove its uniqueness. Let $f\in F$ be such that $f\ne f^*$. Then, there are $R\in\eqdomain$ and $i\in N$ such that $v(R_i)>V^{m+1}(R)$ and  $f_i(R)=\big(1,v(R_i)\big)$. 
        Thus, by \textit{efficiency}, 
        \begin{equation*}
             f^*_i(R)= \big(1,V^{m+1}(R)\big)\mathbin{P_i}\big(1,v(R_i)\big)=f_i^{\mathrm{P}}(R)= f_i(R),
        \end{equation*}
        which implies $f$ is not \textit{agent-welfare-optimal} among $F$.        \qed

    \mbox{}\\
     \textbf{Acknowledgments:} 
    We are deeply grateful to Takehiko Yamato for his insightful guidance and support.
    We also sincerely appreciate helpful comments from Emiko Fukuda, Ryo Kawasaki, Tomomi Matsui, Hiroki Shinozaki, and Akiyoshi Shioura.
    We also thank the participants at Summer Workshop on Game Theory and Experimental Economics 2025 and ERATO Keio seminar. 
     
    \bibliographystyle{apalike} 
       \bibliography{main}

\end{document}